\documentclass[twoside,english,eqsecnum,nofootinbib]{revtex4-2}
\usepackage[T1]{fontenc}
\usepackage[utf8]{inputenc}
\usepackage[a4paper]{geometry}
\usepackage[active]{srcltx}
\usepackage{color}
\usepackage{babel}
\usepackage{bm}
\usepackage{amsmath}
\usepackage{amssymb}
\usepackage[unicode=true,pdfusetitle,
 bookmarks=true,bookmarksnumbered=false,bookmarksopen=false,
 breaklinks=false,pdfborder={0 0 1},backref=false,colorlinks=true]
 {hyperref}

\begin{document}
\title{Spatial covariant gravity with two degrees of freedom in the presence of an auxiliary scalar field: perturbation analysis}
\author{Zhi-Chao Wang}
\author{Xian Gao}
\email{gaoxian@mail.sysu.edu.cn}

\affiliation{School of Physics and Astronomy, Sun Yat-sen University, Zhuhai 519082,
China}

\begin{abstract}
We investigate a class of gravity theories respecting only spatial covariance, termed spatially covariant gravity, in the presence of an auxiliary scalar field. We examine the conditions on the Lagrangian required to eliminate scalar degrees of freedom, allowing only two tensorial degrees of freedom to propagate. Instead of strict constraint analysis, in this work, we employ the perturbation method and focus on the necessary conditions to evade the scalar mode at linear order in perturbations around a cosmological background. Starting from a general action and solving the auxiliary perturbation variables in terms of the would-be dynamical scalar mode, we derive the condition to remove its kinetic term, thus ensuring that no scalar mode propagates. As an application of the general condition, we study a polynomial-type Lagrangian as a concrete example, in which all the monomials are spatially covariant scalars containing two derivatives. We find that the auxiliary scalar field plays a nontrivial role, and new terms in the Lagrangian are allowed. Our analysis sheds light on constructing gravity theories with two degrees of freedom in the extended framework of spatially covariant gravity.
\end{abstract}
\maketitle

\section{Introduction}

Recently, there has been increasing interest in building gravitational theories that differ from general relativity (GR) while propagating the same dynamical degrees of freedom of GR. These degrees of freedom arise as tensorial gravitational waves in a homogeneous and isotropic background, which we thus dub the two tensorial degrees of freedom (TTDOFs). Apparently, it seems impossible since Lovelock's theorem \citep{Lovelock:1971yv,Lovelock:1972vz} claims that GR is the unique metric theory that respects spacetime diffeomorphism and possesses second-order equations of motion in 4 dimensions, which guarantee the propagation of only TTDOFs. Nevertheless, several theories that propagate only TTDOFs have been proposed by bypassing the conditions of Lovelock's theorem.

Theories of TTDOFs can be traced back to the so-called Cuscuton theory \citep{Afshordi:2006ad}, which was extensively studied \citep{Afshordi:2007yx,Mylova:2023ddj,Boruah:2017tvg,Boruah:2018pvq,Bhattacharyya:2016mah,Quintin:2019orx,Bartolo:2021wpt,HosseiniMansoori:2022xnq,Channuie:2023ddv,Maeda:2022ozc,Kohri:2022vst,Panpanich:2021lsd,Lima:2023str} and was further extended in the framework of higher derivative scalar-tensor theory \citep{Iyonaga:2018vnu,Iyonaga:2020bmm}. Previously, such a kind of theory has also been discussed in a subclass of Ho\v{r}ava gravity \citep{Zhu:2011xe,Zhu:2011yu} (see also \citep{Chagoya:2018yna}), with which the relation of Cuscuton has been analyzed \citep{Afshordi:2009tt}. Symmetries behind these theories, including the relation between the TTDOFs (i.e., the transverse and traceless graviton) and the spacetime symmetry \citep{Khoury:2011ay,Khoury:2014sea}, as well as the ``scalarless'' symmetry \citep{Chagoya:2016inc,Tasinato:2020fni}, have also been investigated\footnote{Another attempt was proposed in \citep{Glavan:2019inb} by taking the $D\rightarrow4$ limit of Einstein-Gauss-Bonnet gravity in higher dimensions, which yields an arguable theory propagating only the massless graviton.}.

An alternative framework of theories propagating only two degrees of freedom is dubbed minimally modified gravity (MMG) \citep{Lin:2017oow}. The so-called type-II MMG, which differs from GR in the vacuum, is of particular importance \citep{Aoki:2018brq,Aoki:2018zcv}, which includes the original Cuscuton theory, the minimal theory of massive gravity \citep{DeFelice:2015hla,DeFelice:2015moy} (see also \citep{Bolis:2018vzs,DeFelice:2018vza,DeFelice:2021trp,DeFelice:2023bwq}), as well as the aforementioned $4D$ Einstein-Gauss-Bonnet gravity as special cases. The MMG has been studied extensively \citep{Mukohyama:2019unx,DeFelice:2020eju,DeFelice:2020cpt,Aoki:2020oqc,Pookkillath:2021gdp,DeFelice:2020ecp,Aoki:2021zuy,DeFelice:2020onz,DeFelice:2020prd,DeFelice:2021xps,DeFelice:2022uxv,Jalali:2023wqh,Carballo-Rubio:2018czn,Sangtawee:2021mhz,Ganz:2022iiv,Akarsu:2024qsi}, and has been generalized in the phase space by introducing auxiliary constraints \citep{Yao:2020tur,Yao:2023qjd}.

A general framework of spatially covariant gravity (SCG) theories that propagate TTDOFs with an extra scalar mode was proposed in \citep{Gao:2014soa,Gao:2014fra}, and was further generalized by introducing a dynamical lapse function \citep{Gao:2018znj,Gao:2019lpz} and with nonmetricity \citep{Yu:2024drx}. The SCG can be viewed as an alternative approach to generalizing the scalar-tensor theories \citep{Gao:2020juc,Gao:2020yzr,Gao:2020qxy,Hu:2021bbo}, and includes Ho\v{r}ava gravity \citep{Horava:2008ih,Horava:2009uw,Blas:2009yd}, the effective field theory of inflation/dark energy \citep{Creminelli:2006xe,Cheung:2007st,Creminelli:2008wc,Gubitosi:2012hu,Bloomfield:2012ff} as special cases. The SCG has been applied in the study of cosmology and gravitational waves \citep{Fujita:2015ymn,Gao:2019liu,Zhu:2022dfq,Zhu:2022uoq,Zhu:2023wci}. Within the framework of SCG, conditions on the Lagrangian such that only the TTDOFs are propagating were explored in \citep{Gao:2019twq,Hu:2021yaq} and with a dynamical lapse function in \citep{Lin:2020nro}. Due to the complexity of the conditions, a systematic construction of the TTDOF Lagrangians is still to be performed. Nevertheless, a concrete Lagrangian was found in \citep{Gao:2019twq}, which has been applied in the study of cosmology \citep{Iyonaga:2021yfv,Hiramatsu:2022ahs,Saito:2023bhn,Bartolo:2021wpt,Chakraborty:2023jek}.

Instead of solving the complicated variation-differential equations for the TTDOF conditions in \citep{Gao:2019twq}, an alternative and more straightforward approach is the perturbation method. The idea is to focus on the kinetic term of the would-be dynamical scalar mode. By eliminating its kinetic term order by order in a perturbative expansion around some (typically the cosmological) background, one can fix terms in the Lagrangian order by order. In particular, such a perturbative approach will stop at some finite order since there is a finite number of TTDOF conditions. This perturbation method has been used in \citep{Iyonaga:2018vnu} to extend the Cuscuton theory in the framework of higher derivative scalar-tensor theory. In the framework of SCG, it has also been successfully applied in \citep{Gao:2019lpz} to derive the Lagrangian with a single scalar mode in the presence of a dynamical lapse function, and in \citep{Hu:2021yaq} to build the Lagrangian without any scalar mode, i.e., propagating only TTDOFs.

In this work, we will apply the perturbation method to the SCG in the presence of an auxiliary scalar field and determine the conditions such that the theory propagates only TTDOFs. The idea of SCG with an auxiliary scalar field first arises in \citep{Gao:2018izs}, in which the scalar field was assumed to possess a spacelike gradient and thus loses its kinetic term in the so-called ``spatial gauge''. In this work, we release this assumption and simply treat it as an auxiliary field. In the presence of an auxiliary scalar field, one may have much more room to tune the theory to satisfy the TTDOF conditions and may expect to find a novel class of TTDOF theories. This paper is thus devoted to this issue.

This paper is organized as follows. In Sec. \ref{sec:model}, we describe our model of spatially covariant gravity with an auxiliary scalar field and present the scalar perturbations around a cosmological background. In Sec. \ref{sec:degen}, we derive the quadratic action for the scalar perturbations. By solving the auxiliary perturbation variables and focusing on the kinetic term of the unwanted scalar mode, we derive the degeneracy condition to evade the scalar mode. In Sec. \ref{sec:d2}, we use the polynomial-type Lagrangian of $d=2$ with $d$ being the total number of derivatives as a concrete example and derive the explicit Lagrangian that propagates TTDOFs. In Sec. \ref{sec:con}, we summarize our results.

\section{The perturbation method \label{sec:model}}

\subsection{Model and cosmological perturbations}

Our starting point is the general action of spatial covariant gravity coupled with a non-dynamical scalar field \citep{Gao:2018izs}:
\begin{equation}
	S=\int\mathrm{d}t\mathrm{d}^{3}xN\sqrt{h}\mathcal{L}\left(h_{ij},K_{ij},R_{ij},N,\phi,\mathrm{D}_{i}\right),\label{action}
\end{equation}
where $h_{ij}$ is the spatial metric, $N$ is the lapse function, $R_{ij}$ is the spatial Ricci tensor, $K_{ij}$ is the extrinsic curvature defined by
\begin{equation}
	K_{ij}=\frac{1}{2N}\left(\partial_{t}h_{ij}-\mathrm{D}_{i}N_{j}-\mathrm{D}_{j}N_{i}\right),
\end{equation}
with $N_{i}$ as the shift vector and $\mathrm{D}_{i}$ as the spatial derivative adapted to $h_{ij}$. The scalar field $\phi$ has no kinetic term and thus acts as an auxiliary variable. We emphasize that although this auxiliary scalar field was originally motivated from a scalar field with a spacelike gradient in the so-called spatial gauge, here (and also in the analysis in \citep{Gao:2018izs}) we do not make such an assumption and merely treat it as an auxiliary field, i.e., without any time derivative terms. It has been shown in \citep{Gao:2018izs} that generally the theory (\ref{action}) propagates 3 degrees of freedom, i.e., 2 tensorial and 1 scalar degrees of freedom. The purpose of this work is to find the condition on the Lagrangian in order to eliminate the scalar mode.

The most strict approach to analyzing the number of degrees of freedom is the constraint analysis, either in the Hamiltonian or in the Lagrangian formalism. In the absence of the auxiliary scalar field $\phi$, the exact conditions for the TTDOFs have been found in \citep{Gao:2019twq}. These conditions are variation-differential equations for the Lagrangian, which are difficult (if not impossible) to solve to give the concrete Lagrangian systematically. To address this problem, a perturbation method was employed in \citep{Hu:2021yaq} (and previously in \citep{Gao:2019lpz} with a dynamical lapse function). The advantage of the perturbation method is that one can determine the conditions on the Lagrangian order by order in perturbations, which by themselves are much simpler than the fully nonlinear conditions.

In the following, we will use this perturbation method to determine the conditions on the action (\ref{action}) such that no scalar mode propagates. To be precise, our task is to study the perturbations of the action (\ref{action}) around a cosmological background and eliminate the scalar modes at the linear order in perturbations. The difference of this work from \citep{Hu:2021yaq, Gao:2019lpz} is that here we will first perform the perturbation analysis on the general form of Lagrangian instead of a specific Lagrangian (e.g., in the polynomial form). Therefore, the resulting conditions are generic and can be applied to more general Lagrangians.

For our purpose of eliminating the unwanted scalar degree of freedom, we focus solely on the scalar perturbations. We consider perturbations around a Friedmann-Robertson-Walker background. After fixing the residual gauge freedom of spatial diffeomorphism, the Arnowitt-Deser-Misner (ADM) variables are parameterized as follows:
\begin{eqnarray}
	N & = & \bar{N}e^{A},\\
	N_{i} & = & \bar{N}a\partial_{i}B,\\
	h_{ij} & = & a^{2}e^{2\zeta}\delta_{ij},
\end{eqnarray}
where $a$ is the scale factor. Here, $A$, $B$, and $\zeta$ are scalar perturbations of the metric. The auxiliary scalar field is perturbed as usual:
\begin{equation}
	\phi = \bar{\phi} + \delta\phi.
\end{equation}
All perturbation variables are set to zero at the background level, i.e., $A=B=\zeta=\delta\phi=0$. $\bar{N}$ and $\bar{\phi}$ are the background values of the lapse function and scalar field, respectively, which are functions of time. Note that due to the loss of time reparametrization invariance, we cannot fix the background value of the lapse function $\bar{N}$ to unity. In this paper, we use the same shorthand as in \citep{Fujita:2015ymn}:
\begin{equation}
	\dot{X} = \frac{1}{\bar{N}}\frac{\partial X}{\partial t},
\end{equation}
and
\begin{equation}
	f'=\bar{N}\left.\frac{\partial f}{\partial N}\right|_{N=\bar{N}},\quad f''=\bar{N}^{2}\left.\frac{\partial^{2}f}{\partial N^{2}}\right|_{N=\bar{N}},
\end{equation}
etc.

To obtain the action up to the second order in perturbations, we first
expand all the quantities in the following form
\begin{equation}
Q=Q^{(0)}+\delta Q^{(1)}+\delta Q^{(2)},
\end{equation}
where $Q^{(0)}$ represents the background value of any quantity $Q$,
and $\delta Q^{(n)}$ represents the $n$-th order perturbation of
$Q$. Precisely, we have
\begin{equation}
\phi^{(0)}=\bar{\phi},\quad\delta\phi^{(1)}=\delta\phi,\label{dltphi}
\end{equation}
\begin{equation}
N^{(0)}=\bar{N},\quad\delta N^{(1)}=\bar{N}A,\quad\delta N^{(2)}=\frac{1}{2}\bar{N}A^{2},
\end{equation}
\begin{equation}
N^{i(0)}=0,\quad\delta N^{i(1)}=\bar{N}a^{-1}\partial^{i}B,
\end{equation}
\begin{equation}
h_{ij}^{(0)}=a^{2}\delta_{ij},\quad\delta h_{ij}^{(1)}=2\zeta a^{2}\delta_{ij},\quad\delta h_{ij}^{(2)}=2\zeta^{2}a^{2}\delta_{ij},
\end{equation}
\begin{equation}
K_{ij}^{(0)}=a\dot{a}\delta_{ij},\quad\delta K_{ij}^{(1)}=a\dot{a}\delta_{ij}\left(-A+2\zeta\right)+a^{2}\delta_{ij}\dot{\zeta}-a\partial_{i}\partial_{j}B,
\end{equation}
\begin{eqnarray}
\delta K_{ij}^{(2)} & = & a\dot{a}\delta_{ij}\left(-2\zeta A+\frac{1}{2}A^{2}+2\zeta^{2}\right)+a^{2}\delta_{ij}\dot{\zeta}\left(2\zeta-A\right)\nonumber \\
 &  & +aA\partial_{i}\partial_{j}B+a\left(\partial_{i}\zeta\partial_{j}B+\partial_{j}\zeta\partial_{i}B-\delta_{ij}\partial^{l}B\partial_{l}\zeta\right),
\end{eqnarray}
\begin{equation}
R_{ij}^{(0)}=0,\quad\delta R_{ij}^{(1)}=-\partial_{i}\partial_{j}\zeta-\delta_{ij}\partial^{2}\zeta,
\end{equation}
and
\begin{equation}
\delta R_{ij}^{(2)}=\partial_{i}\zeta\partial_{j}\zeta-\delta_{ij}\partial_{l}\zeta\partial^{l}\zeta.\label{dltRij2}
\end{equation}
In the above and in what follows, spatial indices are raised and lowered by $\delta^{ij}$ or $\delta_{ij}$, respectively. It should be noted that only linear-order terms are considered for the perturbations of the auxiliary scalar field $\phi$ and shift vector $N^{i}$. With the above results, we can proceed to calculate the perturbations of the action in the subsequent sections.

\subsection{Background equations of motion}

We first derive the background equations of motion, which can be used to simplify calculations in higher orders. To this end, we need to obtain the linear-order action $S_{1}$:
\begin{equation}
S_{1}=\int\mathrm{d}t\mathrm{d}^{3}x\left.\frac{\delta S}{\delta\Phi_{I}}\right|_{\mathrm{bg}}\delta\Phi_{I}^{(1)},
\end{equation}
where 
\begin{equation}
\Phi_{I}\equiv\left\{ h_{ij},K_{ij},R_{ij},N,\phi\right\} 
\end{equation}
stands for the basic quantities in our model (\ref{action}), and
``$Q|_{\mathrm{bg}}$'' denotes the background value of any quantity
$Q$. We have
\begin{eqnarray}
S_{1} & = & \int\mathrm{d}t\mathrm{d}^{3}x\left[\bar{N}a^{3}\left.\frac{\delta\mathcal{L}}{\delta K_{ij}}\right|_{\mathrm{bg}}\delta K_{ij}^{(1)}+\left(\bar{N}\frac{1}{2}a\delta^{ij}\bar{\mathcal{L}}+\bar{N}a^{3}\left.\frac{\delta\mathcal{L}}{\delta h_{ij}}\right|_{\mathrm{bg}}\right)\delta h_{ij}^{(1)}+\bar{N}a^{3}\left.\frac{\delta\mathcal{L}}{\delta\phi}\right|_{\mathrm{bg}}\delta\phi\right.\nonumber \\
 & \quad & \left.+\left(a^{3}\bar{\mathcal{L}}+\bar{N}a^{3}\left.\frac{\delta\mathcal{L}}{\delta N}\right|_{\mathrm{bg}}\right)\delta N^{(1)}+\bar{N}a^{3}\left.\frac{\delta\mathcal{L}}{\delta R_{ij}}\right|_{\mathrm{bg}}\delta R_{ij}^{(1)}\right],\label{S1}
\end{eqnarray}
where $\bar{\mathcal{L}}$ represents the background Lagrangian density.
We emphasize that in (\ref{S1}) (and in what follows), when making the variation, $h_{ij}$, $K_{ij}$, $R_{ij}$, $N$, and $\phi$ are treated as independent variables\footnote{For example, although $K_{ij}$, $R_{ij}$, etc., contain $h_{ij}$ implicitly, when evaluating $\frac{\delta\mathcal{L}}{\delta h_{ij}}$, only the explicit functional dependence of $\mathcal{L}$ on $h_{ij}$ will be taken into account.}. Note that at the background level, $N=\bar{N}$, $N_{i}=0$, $h_{ij}=a^{2}\delta_{ij}$, and $\phi=\bar{\phi}$. Plugging (\ref{dltphi})-(\ref{dltRij2}) into (\ref{S1}), after some simplifications, we get
\begin{equation}
S_{1}[A,\zeta,\delta\phi]=\int\mathrm{d}t\mathrm{d}^{3}x\bar{N}a^{3}\left(\mathcal{E}_{A}A+\mathcal{E}_{\zeta}\zeta+\mathcal{E}_{\delta\phi}\delta\phi\right),\label{eq:24}
\end{equation}
where we define
\begin{equation}
\mathcal{E}_{A}=\bar{\mathcal{L}}+\bar{N}\left.\frac{\delta\mathcal{L}}{\delta N}\right|_{\mathrm{bg}}-a^{2}H\delta_{ij}\left.\frac{\delta\mathcal{L}}{\delta K_{ij}}\right|_{\mathrm{bg}},
\end{equation}
with $H\equiv\dot{a}/a\equiv\partial_{t}a/(a\bar{N})$,
\begin{equation}
\mathcal{E}_{\zeta}=3\bar{\mathcal{L}}+a^{2}\delta_{ij}\left[2\left.\frac{\delta\mathcal{L}}{\delta h_{ij}}\right|_{\mathrm{bg}}-3H\left.\frac{\delta\mathcal{L}}{\delta K_{ij}}\right|_{\mathrm{bg}}-\frac{1}{\bar{N}}\frac{\partial}{\partial t}\left(\left.\frac{\delta\mathcal{L}}{\delta K_{ij}}\right|_{\mathrm{bg}}\right)-\frac{\dot{\bar{N}}}{\bar{N}}\left.\frac{\delta\mathcal{L}}{\delta K_{ij}}\right|_{\mathrm{bg}}\right],
\end{equation}
and
\begin{equation}
\mathcal{E}_{\delta\phi}=\left.\frac{\delta\mathcal{L}}{\delta\phi}\right|_{\mathrm{bg}}.
\end{equation}
Requiring the linear variation of the background action to be vanishing, i.e., $S_{1}[A,\zeta,\delta\phi]=0$, yields:
\begin{equation}
	\mathcal{E}_{A}=0,\quad\mathcal{E}_{\zeta}=0,\quad\mathcal{E}_{\delta\phi}=0,\label{bgeom}
\end{equation}
which are the background equations of motion. At this point, note that since SCG explicitly breaks the time diffeomorphism, all three of these equations are independent. This differs from the case of generally covariant theories, in which only two of the three equations are independent due to the time diffeomorphism.

\section{Quadratic action and the degeneracy condition \label{sec:degen}}

The quadratic action for the perturbations receives contributions from both first-order and second-order quantities. Its general form is given by:
\begin{equation}
S_{2}[A,B,\zeta,\delta\phi]=\int\mathrm{d}t\mathrm{d}^{3}x\left(\left.\frac{\delta S}{\delta\Phi_{I}}\right|_{\mathrm{bg}}\delta\Phi_{I}^{(2)}+\frac{1}{2}\left.\frac{\delta^{2}S}{\delta\Phi_{I}\delta\Phi_{J}}\right|_{\mathrm{bg}}\delta\Phi_{I}^{(1)}\delta\Phi_{J}^{(1)}\right),
\end{equation}
where (from now on we suppress the subscript ``bg'' for simplicity)
\begin{eqnarray}
\int\mathrm{d}t\mathrm{d}^{3}x\frac{\delta S}{\delta\Phi_{I}}\delta\Phi_{I}^{(2)} & = & \int\mathrm{d}t\mathrm{d}^{3}x\left[\frac{\delta\left(N\sqrt{h}\mathcal{L}\right)}{\delta K_{ij}}\delta K_{ij}^{(2)}+\frac{\delta\left(N\sqrt{h}\mathcal{L}\right)}{\delta h_{ij}}\delta h_{ij}^{(2)}\right.\nonumber \\
 &  & \qquad\left.+\frac{\delta\left(N\sqrt{h}\mathcal{L}\right)}{\delta N}\delta N^{(2)}+\frac{\delta\left(N\sqrt{h}\mathcal{L}\right)}{\delta R_{ij}}\delta R_{ij}^{(2)}\right],\label{S2_1}
\end{eqnarray}
and
\begin{eqnarray}
 &  & \int\mathrm{d}t\mathrm{d}^{3}x\frac{\delta^{2}S}{\delta\Phi_{I}\delta\Phi_{J}}\delta\Phi_{I}^{(1)}\delta\Phi_{J}^{(1)}\nonumber \\
 & = & \int\mathrm{d}t\mathrm{d}^{3}x\Bigg[2\frac{\delta^{2}\left(N\sqrt{h}\mathcal{L}\right)}{\delta h_{mn}\delta K_{ij}}\delta h_{mn}^{(1)}\delta K_{ij}^{(1)}+\frac{\delta^{2}\left(N\sqrt{h}\mathcal{L}\right)}{\delta h_{mn}\delta h_{ij}}\delta h_{mn}^{(1)}\delta h_{ij}^{(1)}+\frac{\delta^{2}\left(N\sqrt{h}\mathcal{L}\right)}{\delta K_{mn}\delta K_{ij}}\delta K_{mn}^{(1)}\delta K_{ij}^{(1)}\nonumber \\
 &  & +2\frac{\delta^{2}\left(N\sqrt{h}\mathcal{L}\right)}{\delta\phi\delta K_{ij}}\delta\phi\delta K_{ij}^{(1)}+2\frac{\delta^{2}\left(N\sqrt{h}\mathcal{L}\right)}{\delta N\delta K_{ij}}\delta N^{(1)}\delta K_{ij}^{(1)}+2\frac{\delta^{2}\left(N\sqrt{h}\mathcal{L}\right)}{\delta\phi\delta h_{ij}}\delta\phi\delta h_{ij}^{(1)}\nonumber \\
 &  & +2\frac{\delta^{2}\left(N\sqrt{h}\mathcal{L}\right)}{\delta N\delta h_{ij}}\delta N^{(1)}\delta h_{ij}^{(1)}+2\frac{\delta^{2}\left(N\sqrt{h}\mathcal{L}\right)}{\delta\phi\delta N}\delta\phi\delta N^{(1)}+\frac{\delta^{2}\left(N\sqrt{h}\mathcal{L}\right)}{\delta\phi^{2}}\delta\phi\delta\phi\nonumber \\
 &  & +\frac{\delta^{2}\left(N\sqrt{h}\mathcal{L}\right)}{\delta N^{2}}\delta N^{(1)}\delta N^{(1)}+2\frac{\delta^{2}\left(N\sqrt{h}\mathcal{L}\right)}{\delta h_{mn}\delta R_{ij}}\delta h_{mn}^{(1)}\delta R_{ij}^{(1)}+2\frac{\delta^{2}\left(N\sqrt{h}\mathcal{L}\right)}{\delta K_{mn}\delta R_{ij}}\delta K_{mn}^{(1)}\delta R_{ij}^{(1)}\nonumber \\
 &  & +2\frac{\delta^{2}\left(N\sqrt{h}\mathcal{L}\right)}{\delta N\delta R_{ij}}\delta N^{(1)}\delta R_{ij}^{(1)}+\frac{\delta^{2}\left(N\sqrt{h}\mathcal{L}\right)}{\delta R_{mn}\delta R_{ij}}\delta R_{mn}^{(1)}\delta R_{ij}^{(1)}+2\frac{\delta^{2}\left(N\sqrt{h}\mathcal{L}\right)}{\delta\phi\delta R_{ij}}\delta\phi\delta R_{ij}^{(1)}\Bigg].\label{S2_2}
\end{eqnarray}

The quadratic action for the scalar perturbations takes the general form
\begin{eqnarray}
S_{2}\left[A,B,\zeta,\delta\phi\right] & = & \int\mathrm{d}t\mathrm{d}^{3}x\bar{N}a^{3}\left(A\hat{\mathcal{O}}_{AA}A+B\hat{\mathcal{O}}_{BB}B+\zeta\hat{\mathcal{O}}_{\zeta\zeta}\zeta+\delta\phi\hat{\mathcal{O}}_{\phi\phi}\delta\phi+A\hat{\mathcal{O}}_{AB}B+A\hat{\mathcal{O}}_{A\zeta}\zeta+A\hat{\mathcal{O}}_{A\phi}\delta\phi\right.\nonumber \\
 &  & \left.+\delta\phi\hat{\mathcal{O}}_{\phi\zeta}\zeta+\delta\phi\hat{\mathcal{O}}_{\phi B}B+\zeta\hat{\mathcal{O}}_{\zeta B}B+\dot{\zeta}\hat{\mathcal{O}}_{\dot{\zeta}A}A+\dot{\zeta}\hat{\mathcal{O}}_{\dot{\zeta}B}B+\dot{\zeta}\hat{\mathcal{O}}_{\dot{\zeta}\zeta}\zeta+\dot{\zeta}\hat{\mathcal{O}}_{\dot{\zeta}\phi}\delta\phi+\dot{\zeta}\hat{\mathcal{O}}_{\dot{\zeta}\dot{\zeta}}\dot{\zeta}\right),\label{S2}
\end{eqnarray}
where $\hat{\mathcal{O}}_{XY}$ represents the operator involving perturbation variables $X$ and $Y$ (e.g., $\hat{\mathcal{O}}_{A\phi}$ represents the operator involving $A$ and $\delta\phi$). Generally, $\hat{\mathcal{O}}_{XY}$'s are time-dependent and may contain spatial derivatives. By plugging (\ref{dltphi})-(\ref{dltRij2}) into (\ref{S2_1}) and (\ref{S2_2}) and after some tedious manipulations, we get:
\begin{equation}
\hat{\mathcal{O}}_{AA}=\frac{1}{2}\bar{\mathcal{L}}-\frac{1}{2}a\dot{a}\delta_{ij}\frac{\delta\mathcal{L}}{\delta K_{ij}}+\frac{1}{2}a^{2}\dot{a}^{2}\delta_{mn}\delta_{ij}\frac{\delta^{2}\mathcal{L}}{\delta K_{mn}\delta K_{ij}}+\frac{3}{2}\bar{N}\frac{\delta\mathcal{L}}{\delta N}-\bar{N}a\dot{a}\delta_{ij}\frac{\delta^{2}\mathcal{L}}{\delta N\delta K_{ij}}+\frac{1}{2}\bar{N}^{2}\frac{\delta^{2}\mathcal{L}}{\delta N^{2}},\label{OAA_gen}
\end{equation}
\begin{equation}
\hat{\mathcal{O}}_{BB}=\frac{1}{2}a^{2}\frac{\delta^{2}\mathcal{L}}{\delta K_{mn}\delta K_{ij}}\partial_{m}\partial_{n}\partial_{i}\partial_{j},
\end{equation}
\begin{eqnarray}
\hat{\mathcal{O}}_{\zeta\zeta} & = & \frac{9}{2}\bar{\mathcal{L}}+8a^{2}\delta_{ij}\frac{\delta\mathcal{L}}{\delta h_{ij}}+2a^{4}\delta_{mn}\delta_{ij}\frac{\delta^{2}\mathcal{L}}{\delta h_{mn}\delta h_{ij}}+8a\dot{a}\delta_{ij}\frac{\delta\mathcal{L}}{\delta K_{ij}}\nonumber \\
 &  & +4a^{3}\dot{a}\delta_{ij}\delta_{mn}\frac{\delta^{2}\mathcal{L}}{\delta h_{mn}\delta K_{ij}}+2a^{2}\dot{a}^{2}\delta_{mn}\delta_{ij}\frac{\delta^{2}\mathcal{L}}{\delta K_{mn}\delta K_{ij}}\nonumber \\
 &  & +\frac{1}{2}\frac{\delta^{2}\mathcal{L}}{\delta R_{mn}\delta R_{ij}}\left(\partial_{i}\partial_{j}\partial_{m}\partial_{n}+2\delta_{ij}\partial_{m}\partial_{n}\partial^{2}+\delta_{ij}\delta_{mn}\partial^{4}\right)-2a\dot{a}\frac{\delta^{2}\mathcal{L}}{\delta K_{mn}\delta R_{ij}}\delta_{mn}\partial_{i}\partial_{j}\nonumber \\
 &  & -2a^{2}\delta_{mn}\frac{\delta^{2}\mathcal{L}}{\delta h_{mn}\delta R_{ij}}\left(\partial_{i}\partial_{j}+\delta_{ij}\partial^{2}\right)-2\frac{\delta\mathcal{L}}{\delta R_{ij}}\left(2\partial_{i}\partial_{j}+\delta_{ij}\partial^{2}\right),
\end{eqnarray}
\begin{equation}
\hat{\mathcal{O}}_{\phi\phi}=\frac{1}{2}\frac{\delta^{2}\mathcal{L}}{\delta\phi^{2}},
\end{equation}
\begin{equation}
\hat{\mathcal{O}}_{AB}=\left(-\bar{N}a\frac{\delta^{2}\mathcal{L}}{\delta N\delta K_{ij}}+a^{2}\dot{a}\delta_{mn}\frac{\delta^{2}\mathcal{L}}{\delta K_{mn}\delta K_{ij}}\right)\partial_{i}\partial_{j},
\end{equation}
\begin{eqnarray}
\hat{\mathcal{O}}_{A\zeta} & = & 3\bar{\mathcal{L}}+2a^{2}\delta_{ij}\frac{\delta\mathcal{L}}{\delta h_{ij}}+3\bar{N}\frac{\delta\mathcal{L}}{\delta N}+2\bar{N}a^{2}\delta_{ij}\frac{\delta^{2}\mathcal{L}}{\delta N\delta h_{ij}}-3a\dot{a}\delta_{ij}\frac{\delta\mathcal{L}}{\delta K_{ij}}\nonumber \\
 &  & -2a^{3}\dot{a}\delta_{ij}\delta_{mn}\frac{\delta^{2}\mathcal{L}}{\delta h_{mn}\delta K_{ij}}+2\bar{N}a\dot{a}\delta_{ij}\frac{\delta^{2}\mathcal{L}}{\delta N\delta K_{ij}}-2a^{2}\dot{a}^{2}\delta_{mn}\delta_{ij}\frac{\delta^{2}\mathcal{L}}{\delta K_{mn}\delta K_{ij}}\nonumber \\
 &  & -\left(\frac{\delta\mathcal{L}}{\delta R_{ij}}+\bar{N}\frac{\delta^{2}\mathcal{L}}{\delta N\delta R_{ij}}\right)\left(\partial_{i}\partial_{j}+\delta_{ij}\partial^{2}\right)+a\dot{a}\delta_{mn}\frac{\delta^{2}\mathcal{L}}{\delta K_{mn}\delta R_{ij}}\partial_{i}\partial_{j},
\end{eqnarray}
\begin{equation}
\hat{\mathcal{O}}_{A\phi}=\frac{\delta\mathcal{L}}{\delta\phi}+\bar{N}\frac{\delta^{2}\mathcal{L}}{\delta\phi\delta N}-a\dot{a}\delta_{ij}\frac{\delta^{2}\mathcal{L}}{\delta\phi\delta K_{ij}},
\end{equation}
\begin{equation}
\hat{\mathcal{O}}_{\phi\zeta}=3\frac{\delta\mathcal{L}}{\delta\phi}+2a^{2}\delta_{ij}\frac{\delta^{2}\mathcal{L}}{\delta\phi\delta h_{ij}}+2a\dot{a}\delta_{ij}\frac{\delta^{2}\mathcal{L}}{\delta\phi\delta K_{ij}}-\frac{\delta^{2}\mathcal{L}}{\delta\phi\delta R_{ij}}\left(\partial_{i}\partial_{j}+\delta_{ij}\partial^{2}\right),
\end{equation}
\begin{equation}
\hat{\mathcal{O}}_{\phi B}=-a\frac{\delta^{2}\mathcal{L}}{\delta\phi\delta K_{ij}}\partial_{i}\partial_{j},
\end{equation}
\begin{eqnarray}
\hat{\mathcal{O}}_{\zeta B} & = & -5a\frac{\delta\mathcal{L}}{\delta K_{ij}}\partial_{i}\partial_{j}+a\frac{\delta\mathcal{L}}{\delta K_{ij}}\delta_{ij}\partial^{2}-2a^{3}\delta_{mn}\frac{\delta^{2}\mathcal{L}}{\delta h_{mn}\delta K_{ij}}\partial_{i}\partial_{j}-2a^{2}\dot{a}\delta_{mn}\frac{\delta^{2}\mathcal{L}}{\delta K_{mn}\delta K_{ij}}\partial_{i}\partial_{j},\nonumber \\
 &  & +a\frac{\delta^{2}\mathcal{L}}{\delta K_{mn}\delta R_{ij}}\left(\partial_{i}\partial_{j}\partial_{m}\partial_{n}+\delta_{ij}\partial^{2}\partial_{m}\partial_{n}\right),
\end{eqnarray}
\begin{equation}
\hat{\mathcal{O}}_{\dot{\zeta}A}=-a^{3}\dot{a}\delta_{mn}\delta_{ij}\frac{\delta^{2}\mathcal{L}}{\delta K_{mn}\delta K_{ij}}+\bar{N}a^{2}\delta_{ij}\frac{\delta^{2}\mathcal{L}}{\delta N\delta K_{ij}},
\end{equation}
\begin{equation}
\hat{\mathcal{O}}_{\dot{\zeta}B}=-a^{3}\frac{\delta^{2}\mathcal{L}}{\delta K_{mn}\delta K_{ij}}\delta_{mn}\partial_{i}\partial_{j},
\end{equation}
\begin{equation}
\hat{\mathcal{O}}_{\dot{\zeta}\zeta}=5a^{2}\delta_{ij}\frac{\delta\mathcal{L}}{\delta K_{ij}}+2a^{4}\delta_{ij}\delta_{mn}\frac{\delta^{2}\mathcal{L}}{\delta h_{mn}\delta K_{ij}}+2a^{3}\dot{a}\delta_{mn}\delta_{ij}\frac{\delta^{2}\mathcal{L}}{\delta K_{mn}\delta K_{ij}}-a^{2}\delta_{mn}\frac{\delta^{2}\mathcal{L}}{\delta K_{mn}\delta R_{ij}}\left(\partial_{i}\partial_{j}+\delta_{ij}\partial^{2}\right),
\end{equation}
\begin{equation}
\hat{\mathcal{O}}_{\dot{\zeta}\phi}=a^{2}\delta_{ij}\frac{\delta^{2}\mathcal{L}}{\delta\phi\delta K_{ij}},
\end{equation}
and
\begin{equation}
\hat{\mathcal{O}}_{\dot{\zeta}\dot{\zeta}}=\frac{1}{2}a^{4}\frac{\delta^{2}\mathcal{L}}{\delta K_{mn}\delta K_{ij}}\delta_{mn}\delta_{ij}.\label{Ozdzd_gen}
\end{equation}
By deriving the above results, we have made use of integrations by
parts to simplify the expressions. 

By varying the quadratic action (\ref{S2}), one can easily obtain the equations of motion for the perturbation variables $\left\{ A,B,\delta\phi,\zeta\right\}$ as
\begin{eqnarray}
2\hat{\mathcal{O}}_{AA}A+\hat{\mathcal{O}}_{AB}B+\hat{\mathcal{O}}_{A\phi}\delta\phi+\hat{\mathcal{O}}_{A\zeta}\zeta+\hat{\mathcal{O}}_{\dot{\zeta}A}\dot{\zeta} & = & 0,\label{eom1}\\
\hat{\mathcal{O}}_{AB}A+2\hat{\mathcal{O}}_{BB}B+\hat{\mathcal{O}}_{\phi B}\delta\phi+\hat{\mathcal{O}}_{\zeta B}\zeta+\hat{\mathcal{O}}_{\dot{\zeta}B}\dot{\zeta} & = & 0,\\
\hat{\mathcal{O}}_{A\phi}A+\hat{\mathcal{O}}_{\phi B}B+2\hat{\mathcal{O}}_{\phi\phi}\delta\phi+\hat{\mathcal{O}}_{\phi\zeta}\zeta+\hat{\mathcal{O}}_{\dot{\zeta}\phi}\dot{\zeta} & = & 0,\label{eom3}\\
\frac{1}{\bar{N}}\left[2\partial_{t}\left(a^{3}\hat{\mathcal{O}}_{\dot{\zeta}\dot{\zeta}}\dot{\zeta}\right)+\partial_{t}\left(a^{3}\hat{\mathcal{O}}_{\dot{\zeta}\phi}\delta\phi\right)+\partial_{t}\left(a^{3}\hat{\mathcal{O}}_{\dot{\zeta}B}B\right)+\partial_{t}\left(a^{3}\hat{\mathcal{O}}_{\dot{\zeta}A}A\right)+\partial_{t}\left(a^{3}\hat{\mathcal{O}}_{\dot{\zeta}\zeta}\zeta\right)\right]\nonumber \\
-a^{3}\left(2\hat{\mathcal{O}}_{\zeta\zeta}\zeta+\hat{\mathcal{O}}_{A\zeta}A+\hat{\mathcal{O}}_{\phi\zeta}\delta\phi+\hat{\mathcal{O}}_{\zeta B}B\right) & = & 0.\label{eom4}
\end{eqnarray}
From (\ref{eom1})-(\ref{eom4}), it is evident that $A$, $B$, and $\delta\phi$ are all auxiliary variables, as the equations of motion do not contain their first-order time derivative terms. In the case of a non-degenerate coefficient matrix for the perturbation variables $A$, $B$, and $\delta\phi$, i.e.,
\begin{equation}
\varOmega\equiv\det\left(\begin{array}{ccc}
2\hat{\mathcal{O}}_{AA} & \hat{\mathcal{O}}_{AB} & \hat{\mathcal{O}}_{A\phi}\\
\hat{\mathcal{O}}_{AB} & 2\hat{\mathcal{O}}_{BB} & \hat{\mathcal{O}}_{\phi B}\\
\hat{\mathcal{O}}_{A\phi} & \hat{\mathcal{O}}_{\phi B} & 2\hat{\mathcal{O}}_{\phi\phi}
\end{array}\right)\neq0,
\end{equation}
they can be formally solved from their equations of motion (\ref{eom1})-(\ref{eom3}) as
\begin{eqnarray}
A & = & -\frac{1}{2\varOmega}\Big[\Big(\hat{\mathcal{O}}_{A\zeta}\hat{\mathcal{O}}_{\phi B}^{2}-\hat{\mathcal{O}}_{AB}\hat{\mathcal{O}}_{\phi B}\hat{\mathcal{O}}_{\phi\zeta}-\hat{\mathcal{O}}_{A\phi}\hat{\mathcal{O}}_{\phi B}\hat{\mathcal{O}}_{\zeta B}+2\hat{\mathcal{O}}_{A\phi}\hat{\mathcal{O}}_{BB}\hat{\mathcal{O}}_{\phi\zeta}+2\hat{\mathcal{O}}_{AB}\hat{\mathcal{O}}_{\phi\phi}\hat{\mathcal{O}}_{\zeta B}\nonumber \\
 &  & \qquad-4\hat{\mathcal{O}}_{A\zeta}\hat{\mathcal{O}}_{BB}\hat{\mathcal{O}}_{\phi\phi}\Big)\zeta+\Big(\hat{\mathcal{O}}_{\dot{\zeta}A}\hat{\mathcal{O}}_{\phi B}^{2}-\hat{\mathcal{O}}_{AB}\hat{\mathcal{O}}_{\phi B}\hat{\mathcal{O}}_{\dot{\zeta}\phi}-\hat{\mathcal{O}}_{A\phi}\hat{\mathcal{O}}_{\phi B}\hat{\mathcal{O}}_{\dot{\zeta}B}\nonumber \\
 &  & \qquad+2\hat{\mathcal{O}}_{A\phi}\hat{\mathcal{O}}_{BB}\hat{\mathcal{O}}_{\dot{\zeta}\phi}+2\hat{\mathcal{O}}_{AB}\hat{\mathcal{O}}_{\phi\phi}\hat{\mathcal{O}}_{\dot{\zeta}B}-4\hat{\mathcal{O}}_{\dot{\zeta}A}\hat{\mathcal{O}}_{BB}\hat{\mathcal{O}}_{\phi\phi}\Big)\dot{\zeta}\Big],\label{sol_A}
\end{eqnarray}
\begin{eqnarray}
B & = & -\frac{1}{2\varOmega}\Big[\Big(\hat{\mathcal{O}}_{\zeta B}\hat{\mathcal{O}}_{A\phi}^{2}-\hat{\mathcal{O}}_{AB}\hat{\mathcal{O}}_{A\phi}\hat{\mathcal{O}}_{\phi\zeta}-\hat{\mathcal{O}}_{A\phi}\hat{\mathcal{O}}_{\phi B}\hat{\mathcal{O}}_{A\zeta}+2\hat{\mathcal{O}}_{\phi B}\hat{\mathcal{O}}_{AA}\hat{\mathcal{O}}_{\phi\zeta}+2\hat{\mathcal{O}}_{AB}\hat{\mathcal{O}}_{\phi\phi}\hat{\mathcal{O}}_{A\zeta}\nonumber \\
 &  & \qquad-4\hat{\mathcal{O}}_{\zeta B}\hat{\mathcal{O}}_{AA}\hat{\mathcal{O}}_{\phi\phi}\Big)\zeta+\Big(\hat{\mathcal{O}}_{\dot{\zeta}B}\hat{\mathcal{O}}_{A\phi}^{2}-\hat{\mathcal{O}}_{AB}\hat{\mathcal{O}}_{A\phi}\hat{\mathcal{O}}_{\dot{\zeta}\phi}-\hat{\mathcal{O}}_{A\phi}\hat{\mathcal{O}}_{\phi B}\hat{\mathcal{O}}_{\dot{\zeta}A}\nonumber \\
 &  & \qquad+2\hat{\mathcal{O}}_{\phi B}\hat{\mathcal{O}}_{AA}\hat{\mathcal{O}}_{\dot{\zeta}\phi}+2\hat{\mathcal{O}}_{AB}\hat{\mathcal{O}}_{\phi\phi}\hat{\mathcal{O}}_{\dot{\zeta}A}-4\hat{\mathcal{O}}_{\dot{\zeta}B}\hat{\mathcal{O}}_{AA}\hat{\mathcal{O}}_{\phi\phi}\Big)\dot{\zeta}\Big],
\end{eqnarray}
\begin{eqnarray}
\delta\phi & = & -\frac{1}{2\varOmega}\Big[\Big(\hat{\mathcal{O}}_{\phi\zeta}\hat{\mathcal{O}}_{AB}^{2}-\hat{\mathcal{O}}_{AB}\hat{\mathcal{O}}_{A\zeta}\hat{\mathcal{O}}_{\phi B}-\hat{\mathcal{O}}_{AB}\hat{\mathcal{O}}_{A\phi}\hat{\mathcal{O}}_{\zeta B}+2\hat{\mathcal{O}}_{\phi B}\hat{\mathcal{O}}_{AA}\hat{\mathcal{O}}_{\zeta B}+2\hat{\mathcal{O}}_{BB}\hat{\mathcal{O}}_{A\phi}\hat{\mathcal{O}}_{A\zeta}\nonumber \\
 &  & \qquad-4\hat{\mathcal{O}}_{BB}\hat{\mathcal{O}}_{AA}\hat{\mathcal{O}}_{\phi\zeta}\Big)\zeta+\Big(\hat{\mathcal{O}}_{\dot{\zeta}\phi}\hat{\mathcal{O}}_{AB}^{2}-\hat{\mathcal{O}}_{AB}\hat{\mathcal{O}}_{\dot{\zeta}A}\hat{\mathcal{O}}_{\phi B}-\hat{\mathcal{O}}_{AB}\hat{\mathcal{O}}_{A\phi}\hat{\mathcal{O}}_{\dot{\zeta}B}\nonumber \\
 &  & \qquad+2\hat{\mathcal{O}}_{\phi B}\hat{\mathcal{O}}_{AA}\hat{\mathcal{O}}_{\dot{\zeta}B}+2\hat{\mathcal{O}}_{BB}\hat{\mathcal{O}}_{A\phi}\hat{\mathcal{O}}_{\dot{\zeta}A}-4\hat{\mathcal{O}}_{BB}\hat{\mathcal{O}}_{AA}\hat{\mathcal{O}}_{\dot{\zeta}\phi}\Big)\dot{\zeta}\Big],\label{sol_dltphi}
\end{eqnarray}
with
\begin{equation}
\varOmega=\hat{\mathcal{O}}_{A\phi}^{2}\hat{\mathcal{O}}_{BB}+\hat{\mathcal{O}}_{AA}\hat{\mathcal{O}}_{\phi B}^{2}+\hat{\mathcal{O}}_{AB}^{2}\hat{\mathcal{O}}_{\phi\phi}-\hat{\mathcal{O}}_{AB}\hat{\mathcal{O}}_{A\phi}\hat{\mathcal{O}}_{\phi B}-4\hat{\mathcal{O}}_{AA}\hat{\mathcal{O}}_{BB}\hat{\mathcal{O}}_{\phi\phi}.\label{eq:57}
\end{equation}
Although we have made the assumption that $\varOmega\neq0$ to ensure that the auxiliary variables $A$, $B$, and $\delta\phi$ are solvable (at least formally), it is important to note that even if $\varOmega\neq0$ is not satisfied, it is still possible to obtain conditions to eliminate the unwanted scalar degree of freedom. We will show this explicitly through a detailed analysis of a specific example in the next section.

Finally, by plugging the solutions (\ref{sol_A})-(\ref{sol_dltphi}) into the equation of motion for $\zeta$ (\ref{eom4}), we will obtain an equation of motion for the single variable $\zeta$. If no further conditions are assumed, the effective equation of motion for $\zeta$ will contain its second-order time derivative term $\ddot{\zeta}$, indicating a propagating degree of freedom carried by $\zeta$. For our purpose, in order to eliminate the scalar degree of freedom, we have to ensure that $\zeta$ is not propagating, at least at linear order in a cosmological background. To this end, we have to make the coefficient of $\ddot{\zeta}$ vanish. After some manipulations, the coefficient of $\ddot{\zeta}$ is found to be
\begin{eqnarray}
\varDelta & = & 4\hat{\mathcal{O}}_{\dot{\zeta}\dot{\zeta}}\left(\hat{\mathcal{O}}_{A\phi}^{2}\hat{\mathcal{O}}_{BB}+\hat{\mathcal{O}}_{AA}\hat{\mathcal{O}}_{\phi B}^{2}+\hat{\mathcal{O}}_{AB}^{2}\hat{\mathcal{O}}_{\phi\phi}-\hat{\mathcal{O}}_{AB}\hat{\mathcal{O}}_{A\phi}\hat{\mathcal{O}}_{\phi B}-4\hat{\mathcal{O}}_{AA}\hat{\mathcal{O}}_{BB}\hat{\mathcal{O}}_{\phi\phi}\right)\nonumber \\
 &  & -\hat{\mathcal{O}}_{\phi B}^{2}\hat{\mathcal{O}}_{\dot{\zeta}A}^{2}-\hat{\mathcal{O}}_{A\phi}^{2}\hat{\mathcal{O}}_{\dot{\zeta}B}^{2}-\hat{\mathcal{O}}_{AB}^{2}\hat{\mathcal{O}}_{\dot{\zeta}\phi}^{2}+2\hat{\mathcal{O}}_{A\phi}\hat{\mathcal{O}}_{\phi B}\hat{\mathcal{O}}_{\dot{\zeta}A}\hat{\mathcal{O}}_{\dot{\zeta}B}+2\hat{\mathcal{O}}_{AB}\hat{\mathcal{O}}_{\phi B}\hat{\mathcal{O}}_{\dot{\zeta}A}\hat{\mathcal{O}}_{\dot{\zeta}\phi}+2\hat{\mathcal{O}}_{AB}\hat{\mathcal{O}}_{A\phi}\hat{\mathcal{O}}_{\dot{\zeta}B}\hat{\mathcal{O}}_{\dot{\zeta}\phi}\nonumber \\
 &  & +4\hat{\mathcal{O}}_{BB}\hat{\mathcal{O}}_{\phi\phi}\hat{\mathcal{O}}_{\dot{\zeta}A}^{2}+4\hat{\mathcal{O}}_{AA}\hat{\mathcal{O}}_{BB}\hat{\mathcal{O}}_{\dot{\zeta}\phi}^{2}+4\hat{\mathcal{O}}_{AA}\hat{\mathcal{O}}_{\phi\phi}\hat{\mathcal{O}}_{\dot{\zeta}B}^{2}\nonumber \\
 &  & -4\hat{\mathcal{O}}_{AA}\hat{\mathcal{O}}_{\phi B}\hat{\mathcal{O}}_{\dot{\zeta}B}\hat{\mathcal{O}}_{\dot{\zeta}\phi}-4\hat{\mathcal{O}}_{AB}\hat{\mathcal{O}}_{\phi\phi}\hat{\mathcal{O}}_{\dot{\zeta}A}\hat{\mathcal{O}}_{\dot{\zeta}B}-4\hat{\mathcal{O}}_{A\phi}\hat{\mathcal{O}}_{BB}\hat{\mathcal{O}}_{\dot{\zeta}A}\hat{\mathcal{O}}_{\dot{\zeta}\phi}.\label{eq:59}
\end{eqnarray}
Therefore, we have to require that 
\begin{equation}
\varDelta=0,\label{eq:58}
\end{equation}
which is a necessary condition to eliminate the scalar degree of freedom (i.e., the TTDOF conditions), at least at the linear order in perturbations around a cosmological background.

\section{Theory of $d=2$ \label{sec:d2}}

The analysis and conditions obtained in the above section are general but quite formal. In this section, we will apply the condition (\ref{eq:59}) to a concrete model as an example. According to the classification of SCG monomials \citep{Gao:2020juc, Gao:2020yzr}, we consider a polynomial-type Lagrangian built of monomials of $d=2$, where $d$ is the total number of derivatives. Precisely, the action is given by:
\begin{equation}
S_{2}=\int dtd^{3}xN\sqrt{h}\left(\mathcal{L}-\Lambda\right),\label{eq:60}
\end{equation}
with
\begin{equation}
\mathcal{L}=c_{1}K_{ij}K^{ij}+c_{2}K^{2}+c_{3}R+c_{4}a_{i}a^{i}+d_{1}\mathrm{D}_{i}\phi\mathrm{D}^{i}\phi+d_{2}a_{i}\mathrm{D}^{i}\phi,\label{Lag}
\end{equation}
where $c_{i}$ and $d_{i}$ are general functions of $N$ and $\phi$, and $\Lambda$ is the cosmological constant. Additionally, the acceleration $a_{i}$ is defined as $a_{i}=\partial_{i}\ln N$. We make the assumption that both $\phi$ and $N$ are homogeneous and isotropic at the background, i.e., $\bar{\phi}=\bar{\phi}(t)$, $\bar{N}=\bar{N}(t)$. To simplify our calculation, we denote:
\begin{equation}
	\mathcal{L}_{c}=\mathcal{L}-\Lambda.
\end{equation}
We will see that a non-vanishing cosmological constant $\Lambda$, which may be time-dependent, is necessary in order to have a cosmological background solution.

\subsection{The degeneracy condition}

Since both $\phi$ and $N$ depend only on time at the background level, the background values of each quantity in the Lagrangian are given by:
\begin{equation}
\bar{K}_{ij}=a\dot{a}\delta_{ij},\quad\bar{K}^{ij}=Ha^{-2}\delta^{ij},\quad\bar{K}=3H,
\end{equation}
and
\begin{equation}
\bar{R}=0,\quad\bar{a}_{i}=0,\quad\partial_{i}\bar{\phi}=0.
\end{equation}
The background value of the Lagrangian is
\begin{equation}
\bar{\mathcal{L}_{c}}=3H^{2}b-\Lambda,\label{eq:68}
\end{equation}
where we define
\begin{equation}
b=c_{1}+3c_{2}\label{b_def}
\end{equation}
for short. The non-vanishing derivatives of the Lagrangian $\mathcal{L}_{c}$
with respect to various quantities are thus
\begin{equation}
\frac{\delta\mathcal{L}_{c}}{\delta N}=\frac{3}{\bar{N}}H^{2}b',
\end{equation}
\begin{equation}
\frac{\delta\mathcal{L}_{c}}{\delta K_{ij}}=2Ha^{-2}\delta^{ij}b,
\end{equation}
\begin{equation}
\frac{\delta\mathcal{L}_{c}}{\delta\phi}=3H^{2}\frac{\partial b}{\partial\phi},
\end{equation}
\begin{equation}
\frac{\delta^{2}\mathcal{L}_{c}}{\delta N^{2}}=\frac{3}{\bar{N}^{2}}H^{2}b''-\frac{2}{\bar{N}^{2}}c_{4}\partial^{2},
\end{equation}
\begin{equation}
\frac{\delta^{2}\mathcal{L}_{c}}{\delta K_{mn}\delta K_{ij}}=2a^{-4}\left(\frac{1}{2}c_{1}\left(\delta^{im}\delta^{jn}+\delta^{in}\delta^{jm}\right)+c_{2}\delta^{ij}\delta^{mn}\right),
\end{equation}
\begin{equation}
\frac{\delta^{2}\mathcal{L}_{c}}{\delta\phi^{2}}=3H^{2}\frac{\partial^{2}b}{\partial\phi^{2}}-2d_{1}\partial^{2},
\end{equation}
\begin{equation}
\frac{\delta^{2}\mathcal{L}_{c}}{\delta N\delta K_{ij}}=\frac{2}{\bar{N}}Ha^{-2}\delta^{ij}b',
\end{equation}
\begin{equation}
\frac{\delta^{2}\mathcal{L}_{c}}{\delta\phi\delta N}=3H^{2}\frac{\partial^{2}b}{\partial\phi\partial N}-\frac{d_{2}}{\bar{N}}\partial^{2},
\end{equation}
and
\begin{equation}
\frac{\delta^{2}\mathcal{L}_{c}}{\delta\phi\delta K_{ij}}=2Ha^{-2}\delta^{ij}\frac{\partial b}{\partial\phi}.\label{eq:77}
\end{equation}
We do not present derivatives with respect to $R_{ij}$, as they are irrelevant to the degeneracy analysis. According to (\ref{eq:68})-(\ref{eq:77}), the background equations of motion (\ref{bgeom}) are given by:
\begin{equation}
\mathcal{E}_{A}=3H^{2}\left(-b+b'\right)-\Lambda=0,\label{bgeom_A}
\end{equation}
\begin{equation}
\mathcal{E}_{\zeta}=-9H^{2}b-6\dot{H}b-\frac{6H}{\bar{N}}\dot{\bar{N}}b-3\Lambda=0,\label{bgeom_z}
\end{equation}
\begin{equation}
\mathcal{E}_{\delta\phi}=3H^{2}\frac{\partial b}{\partial\phi}=0,\label{bgeom_dp}
\end{equation}
explicitly. From (\ref{bgeom_dp}), it is clear that in order to have a homogeneous and isotropic background, $b$ defined in (\ref{b_def}) must be a function of the lapse function $N$ only. The background equations of motion (\ref{bgeom_A})-(\ref{bgeom_dp}) can greatly simplify the calculation of the quadratic action for the perturbations in the following.

By plugging (\ref{eq:68})-(\ref{eq:77}) into (\ref{OAA_gen})-(\ref{Ozdzd_gen}), we can evaluate the operators $\hat{\mathcal{O}}_{XY}$ that are relevant to eliminating the scalar mode, which are given by:
\begin{equation}
\hat{\mathcal{O}}_{AA}=\frac{3}{2}H^{2}\left(2b-2b'+b''\right)-c_{4}\partial^{2},\label{OAA_OmgN0}
\end{equation}
\begin{equation}
\hat{\mathcal{O}}_{BB}=a^{-2}\left(c_{1}+c_{2}\right)\partial^{4},
\end{equation}
\begin{equation}
\hat{\mathcal{O}}_{\phi\phi}=\frac{3}{2}H^{2}\frac{\partial^{2}b}{\partial\phi^{2}}-d_{1}\partial^{2},
\end{equation}
\begin{equation}
\hat{\mathcal{O}}_{AB}=2Ha^{-1}\left(-b'+b\right)\partial^{2},
\end{equation}
\begin{equation}
\hat{\mathcal{O}}_{A\phi}=3H^{2}\frac{\partial\left(b'-b\right)}{\partial\phi}-d_{2}\partial^{2},
\end{equation}
\begin{equation}
\hat{\mathcal{O}}_{\phi B}=-2Ha^{-1}\frac{\partial b}{\partial\phi}\partial^{2},
\end{equation}
\begin{equation}
\hat{\mathcal{O}}_{\dot{\zeta}A}=6H\left(-b+b'\right),
\end{equation}
\begin{equation}
\hat{\mathcal{O}}_{\dot{\zeta}B}=-2a^{-1}b\partial^{2},
\end{equation}
\begin{equation}
\hat{\mathcal{O}}_{\dot{\zeta}\phi}=6H\frac{\partial b}{\partial\phi},
\end{equation}
and
\begin{equation}
\hat{\mathcal{O}}_{\dot{\zeta}\dot{\zeta}}=3b.\label{Ozdzd_OmgN0}
\end{equation}

Then, by plugging (\ref{OAA_OmgN0})-(\ref{Ozdzd_OmgN0}) into the determinant of the coefficient matrix (\ref{eq:57}), and keeping in mind that $\frac{\partial b}{\partial\phi}=0$ from the background equation of motion (\ref{bgeom_dp}), we find:
\begin{align}
\varOmega & =a^{-2}\left(c_{1}+c_{2}\right)\left(d_{2}^{2}-4d_{1}c_{4}\right)\partial^{8}\nonumber \\
 & \quad+2H^{2}a^{-2}\left[3\left(2b-2b'+b''\right)\left(c_{1}+c_{2}\right)-2\left(-b'+b\right)^{2}\right]d_{1}\partial^{6}.\label{eq:91}
\end{align}
The degeneracy condition (\ref{eq:58}) reads
\begin{eqnarray}
0\equiv\varDelta & = & 8a^{-2}bc_{1}\left(d_{2}^{2}-4c_{4}d_{1}\right)\partial^{8}\nonumber \\
 &  & +48H^{2}a^{-2}d_{1}c_{1}\left(2bb'-2b'^{2}+bb''\right)\partial^{6}.\label{eq:92}
\end{eqnarray}
In deriving (\ref{eq:91}) and (\ref{eq:92}), the background equations of motion (\ref{bgeom_A})-(\ref{bgeom_dp}) have been used to simplify the expressions.

In the following, we will discuss two cases based on whether $\varOmega$ is vanishing or not.

\subsection{Case $\varOmega\protect\neq0$}

When $\varOmega\neq0$, all the auxiliary variables $A$, $B$, $\delta\phi$ are solvable, and thus the degeneracy condition $\varDelta=0$ holds. According to (\ref{eq:92}), in order to eliminate the unwanted scalar degree of freedom, we must require the coefficients of the operators $\partial^{6}$ and $\partial^{8}$ to be vanishing simultaneously, i.e.,
\begin{equation}
	\left(2bb'-2b'^{2}+bb''\right)c_{1}d_{1}=0,\label{eq:93}
\end{equation}
and
\begin{equation}
	bc_{1}\left(d_{2}^{2}-4c_{4}d_{1}\right)=0,\label{eq:94}
\end{equation}
which are two constraints among $c_{1}$, $c_{4}$, $d_{1}$, $d_{2}$, and $b$. Before solving these two equations, let us conduct a brief analysis.

First, (\ref{eq:93}) and (\ref{eq:94}) are trivially satisfied if $b\equiv c_{1}+3c_{2}=0$. However, this conflicts with the background equation of motion (\ref{bgeom_A}), which leads to $\Lambda=0$. As we have mentioned before, a non-vanishing (and actually positive) cosmological constant is required in order to have a cosmological background solution. Therefore, we must require $b\neq0$. Second, (\ref{eq:93}) and (\ref{eq:94}) are also satisfied if $c_{1}=0$. However, although in this paper we focus on the scalar perturbations only, further calculation reveals that only the term $c_{1}K_{ij}K^{ij}$ will contribute to the kinetic term for the tensor perturbations (i.e., the gravitational waves). As a result, a vanishing $c_{1}$ would also eliminate the gravitational waves, which is unacceptable in light of the observation of gravitational waves. Thus, we must also require $c_{1}\neq0$. After taking into account that $b,c_{1}\neq0$, (\ref{eq:94}) holds only if $d_{2}^{2}-4c_{4}d_{1}=0.$ Finally, if $d_{1}=0$, it follows that $d_{2}=0$. From (\ref{eq:91}), it will lead to $\varOmega=0$, which conflicts with our requirement $\varOmega\neq0$. Thus, we must have $d_{1}\neq0$.

Based on the above arguments, the equations (\ref{eq:93})-(\ref{eq:94}) hold only if:
\begin{align}
d_{2}^{2}-4c_{4}d_{1} & =0,\label{eq:95}\\
2bb'-2b'^{2}+bb'' & =0,\label{eq:96}
\end{align}
are satisfied. The general solutions for $c_{4}$ and $b$ to (\ref{eq:95})-(\ref{eq:96}) are
\begin{equation}
c_{4}=\frac{d_{2}^{2}}{4d_{1}},
\end{equation}
\begin{equation}
b=\frac{C_{2}N}{1+C_{1}N},
\end{equation}
where $C_{1}$ and $C_{2}$ are constants. Finally, the Lagrangian containing no dynamical scalar degree of freedom at linear order in a cosmological background is
\begin{equation}
\mathcal{L}=c_{1}\hat{K}_{ij}\hat{K}^{ij}+\frac{1}{3}\frac{C_{2}N}{1+C_{1}N}K^{2}+c_{3}R+\frac{d_{2}^{2}}{4d_{1}}a_{i}a^{i}+d_{1}\mathrm{D}_{i}\phi\mathrm{D}^{i}\phi+d_{2}a_{i}\mathrm{D}^{i}\phi,\label{Lag_OmgN0}
\end{equation}
where $\hat{K}_{ij}=K_{ij}-\frac{1}{3}Kh_{ij}$ is the traceless part of $K_{ij}$, and the coefficient $c_{1}$ is a function of $N$ only, while $c_{3}$, $d_{1}$, and $d_{2}$ are general functions of $N$ and $\phi$.

It is interesting to compare (\ref{Lag_OmgN0}) with the result obtained in \citep{Hu:2021yaq}. In \citep{Hu:2021yaq} (see eq. (61)), only the first three terms in (\ref{Lag_OmgN0}) are present. In particular, the acceleration term $a_{i}a^{i}$ is not allowed. Here, due to the presence of the auxiliary field $\phi$, the acceleration term can be introduced, together with two other terms involving $\mathrm{D}_{i}\phi$. By taking an appropriate limit $d_{1},d_{2}\rightarrow0$, one will return to the result in \citep{Hu:2021yaq}.

\subsection{{\normalsize{}Case $\varOmega=0$}}

If $\varOmega=0$, not all of the auxiliary variables $A$, $B$, $\delta\phi$ are solvable. Nevertheless, we can still determine the conditions on the coefficients in order to eliminate the scalar degree of freedom. First of all, according to (\ref{eq:91}), $\varOmega=0$ implies two equations:
\begin{equation}
	\left(c_{1}+c_{2}\right)\left(d_{2}^{2}-4d_{1}c_{4}\right)=0,\label{Omg0_eq1}
\end{equation}
and
\begin{equation}
	\left[3\left(2b-2b'+b''\right)\left(c_{1}+c_{2}\right)-2\left(-b'+b\right)^{2}\right]d_{1}=0.\label{Omg0_eq2}
\end{equation}
There are several branches of solutions to (\ref{Omg0_eq1}) and (\ref{Omg0_eq2}), which we will discuss below.

\subsubsection{Case 1}

The simplest case is to assume $d_{1}=d_{2}=0$. In this case, the Lagrangian (\ref{Lag}) reduces to:
\begin{equation}
	\mathcal{L}=c_{1}K_{ij}K^{ij}+c_{2}K^{2}+c_{3}R+c_{4}a_{i}a^{i},
\end{equation}
which is just the case that has already been discussed in \citep{Hu:2021yaq}. The Lagrangian that propagates only two tensor degrees of freedom (up to the linear order in perturbations) is given by
\begin{equation}
	\mathcal{L}=c_{1}\hat{K}_{ij}\hat{K}^{ij}+\frac{1}{3}\frac{C_{2}N}{1+C_{1}N}K^{2}+c_{3}R,\label{eq:103}
\end{equation}
where $c_{1}$ is a function of $N$ (recall that $b\equiv c_{1}+3c_{2}$ has no $\phi$-dependence) and $c_{3}$ is a general function of $N$ and $\phi$.

\subsubsection{Case 2}

If $d_{1}\neq0$, from (\ref{Omg0_eq2}) we must have
\begin{equation}
3\left(2b-2b'+b''\right)\left(b-2c_{2}\right)-2\left(-b'+b\right)^{2}=0.\label{eq:105}
\end{equation}
As for (\ref{Omg0_eq1}), we assume $c_{1}+c_{2}\neq0$ and thus it follows that
\begin{equation}
d_{2}^{2}-4d_{1}c_{4}=0.
\end{equation}

Since the background equation of motion (\ref{bgeom_A}) must hold, we require $-b+b'\neq0$ to guarantee a non-vanishing cosmological constant $\Lambda$ and thus a cosmological background solution. Thus, (\ref{eq:105}) indicates:
\begin{equation}
b-2c_{2}\neq0,\quad2b-2b'+b''\neq0.
\end{equation}
Then (\ref{eq:105}) implies that
\begin{equation}
c_{2}=\frac{1}{2}b-\frac{\left(b-b'\right)^{2}}{3\left(2b-2b'+b''\right)},
\end{equation}
and
\begin{equation}
c_{1}=-\frac{1}{2}b+\frac{\left(b-b'\right)^{2}}{2b-2b'+b''},
\end{equation}
where $b$ is a general function of $N$ only. The Lagrangian now turns to be
\begin{equation}
\mathcal{L}=\left[-\frac{1}{2}b+\frac{\left(b-b'\right)^{2}}{2b-2b'+b''}\right]K_{ij}K^{ij}+\left[\frac{1}{2}b-\frac{\left(b-b'\right)^{2}}{3\left(2b-2b'+b''\right)}\right]K^{2}+c_{3}R+c_{4}a_{i}a^{i}+d_{1}\mathrm{D}_{i}\phi\mathrm{D}^{i}\phi+d_{2}a_{i}\mathrm{D}^{i}\phi,
\end{equation}
where $c_{3}$, $c_{4}$, $d_{1}$ and $d_{2}$ are general functions
of $N$ and $\phi$. 

It is not difficult to check that the operators $\hat{\mathcal{O}}_{XY}$ are the same as in (\ref{OAA_OmgN0})-(\ref{Ozdzd_OmgN0}). The equations of motion for the perturbation variables are:
\begin{eqnarray}
2\hat{\mathcal{O}}_{AA}A+\hat{\mathcal{O}}_{AB}B+\hat{\mathcal{O}}_{A\phi}\delta\phi+\hat{\mathcal{O}}_{A\zeta}\zeta+\hat{\mathcal{O}}_{\dot{\zeta}A}\dot{\zeta} & = & 0,\label{eq:111}\\
\hat{\mathcal{O}}_{AB}A+2\hat{\mathcal{O}}_{BB}B+\hat{\mathcal{O}}_{\zeta B}\zeta+\hat{\mathcal{O}}_{\dot{\zeta}B}\dot{\zeta} & = & 0,\\
2\hat{\mathcal{O}}_{\phi\phi}\delta\phi+\hat{\mathcal{O}}_{A\phi}A & = & 0,\label{eq:113}\\
\frac{1}{\bar{N}}\left[2\partial_{t}\left(a^{3}\hat{\mathcal{O}}_{\dot{\zeta}\dot{\zeta}}\dot{\zeta}\right)+\partial_{t}\left(a^{3}\hat{\mathcal{O}}_{\dot{\zeta}B}B\right)+\partial_{t}\left(a^{3}\hat{\mathcal{O}}_{\dot{\zeta}A}A\right)+\partial_{t}\left(a^{3}\hat{\mathcal{O}}_{\dot{\zeta}\zeta}\zeta\right)\right]\nonumber \\
-a^{3}\left(2\hat{\mathcal{O}}_{\zeta\zeta}\zeta+\hat{\mathcal{O}}_{A\zeta}A+\hat{\mathcal{O}}_{\zeta B}B\right) & = & 0.\label{eq:114}
\end{eqnarray}
We rewrite the first three equations in the matrix form:
\begin{equation}
\bm{D}\bm{\alpha}=\bm{\beta},
\end{equation}
where
\begin{equation}
\bm{D}=\left[\begin{array}{ccc}
2\hat{\mathcal{O}}_{AA} & \hat{\mathcal{O}}_{AB} & \hat{\mathcal{O}}_{A\phi}\\
\hat{\mathcal{O}}_{AB} & 2\hat{\mathcal{O}}_{BB} & 0\\
\hat{\mathcal{O}}_{A\phi} & 0 & 2\hat{\mathcal{O}}_{\phi\phi}
\end{array}\right],
\end{equation}
\begin{equation}
\bm{\alpha}=\left[\begin{array}{c}
A\\
B\\
\delta\phi
\end{array}\right],\quad\bm{\beta}=\left[\begin{array}{c}
-\hat{\mathcal{O}}_{A\zeta}\zeta-\hat{\mathcal{O}}_{\dot{\zeta}A}\dot{\zeta}\\
-\hat{\mathcal{O}}_{\zeta B}\zeta-\hat{\mathcal{O}}_{\dot{\zeta}B}\dot{\zeta}\\
0
\end{array}\right].
\end{equation}

The determinant of coefficient matrix $\bm{D}$ of the auxiliary perturbation variables $A$, $B$, and $\delta\phi$ is
\begin{align}
\varOmega & =\det\bm{D}\nonumber \\
 & =\hat{\mathcal{O}}_{A\phi}^{2}\hat{\mathcal{O}}_{BB}+\hat{\mathcal{O}}_{AB}^{2}\hat{\mathcal{O}}_{\phi\phi}-4\hat{\mathcal{O}}_{AA}\hat{\mathcal{O}}_{BB}\hat{\mathcal{O}}_{\phi\phi}=0.
\end{align}
By multiplying the third row of $\bm{D}$ by $\hat{\mathcal{O}}_{A\phi}\hat{\mathcal{O}}_{BB}$, the second row by $\hat{\mathcal{O}}_{AB}\hat{\mathcal{O}}_{\phi\phi}$, and the first row by $-2\hat{\mathcal{O}}_{BB}\hat{\mathcal{O}}_{\phi\phi}$, and adding the third and the second rows to the first row, we arrive at the following matrix:
\begin{equation}
\left(\begin{array}{ccc}
0 & 0 & 0\\
\hat{\mathcal{O}}_{AB} & 2\hat{\mathcal{O}}_{BB} & 0\\
\hat{\mathcal{O}}_{A\phi} & 0 & 2\hat{\mathcal{O}}_{\phi\phi}
\end{array}\right),
\end{equation}
from which it is transparent that $\mathrm{rank}\bm{D}=2<3$. The
augmented matrix of the set of equations is defined as
\begin{equation}
\bm{E}=\left(\begin{array}{cccc}
2\hat{\mathcal{O}}_{AA} & \hat{\mathcal{O}}_{AB} & \hat{\mathcal{O}}_{A\phi} & -\hat{\mathcal{O}}_{A\zeta}\zeta-\hat{\mathcal{O}}_{\dot{\zeta}A}\dot{\zeta}\\
\hat{\mathcal{O}}_{AB} & 2\hat{\mathcal{O}}_{BB} & 0 & -\hat{\mathcal{O}}_{\zeta B}\zeta-\hat{\mathcal{O}}_{\dot{\zeta}B}\dot{\zeta}\\
\hat{\mathcal{O}}_{A\phi} & 0 & 2\hat{\mathcal{O}}_{\phi\phi} & 0
\end{array}\right).
\end{equation}
By performing the same operations to $\bm{D}$, the augmented matrix becomes
\begin{equation}
\left(\begin{array}{cccc}
0 & 0 & 0 & \left(2\hat{\mathcal{O}}_{BB}\hat{\mathcal{O}}_{\phi\phi}\hat{\mathcal{O}}_{A\zeta}-\hat{\mathcal{O}}_{\zeta B}\hat{\mathcal{O}}_{AB}\hat{\mathcal{O}}_{\phi\phi}\right)\zeta+\left(2\hat{\mathcal{O}}_{BB}\hat{\mathcal{O}}_{\phi\phi}\hat{\mathcal{O}}_{\dot{\zeta}A}-\hat{\mathcal{O}}_{\dot{\zeta}B}\hat{\mathcal{O}}_{AB}^{2}\hat{\mathcal{O}}_{\phi\phi}\right)\dot{\zeta}\\
\hat{\mathcal{O}}_{AB} & 2\hat{\mathcal{O}}_{BB} & 0 & -\hat{\mathcal{O}}_{\zeta B}\zeta-\hat{\mathcal{O}}_{\dot{\zeta}B}\dot{\zeta}\\
\hat{\mathcal{O}}_{A\phi} & 0 & 2\hat{\mathcal{O}}_{\phi\phi} & 0
\end{array}\right),
\end{equation}
from which it is clear that the crucial term relevant to our analysis is
\begin{equation}
\Xi\equiv\left(2\hat{\mathcal{O}}_{BB}\hat{\mathcal{O}}_{\phi\phi}\hat{\mathcal{O}}_{A\zeta}-\hat{\mathcal{O}}_{\zeta B}\hat{\mathcal{O}}_{AB}\hat{\mathcal{O}}_{\phi\phi}\right)\zeta+\left(2\hat{\mathcal{O}}_{BB}\hat{\mathcal{O}}_{\phi\phi}\hat{\mathcal{O}}_{\dot{\zeta}A}-\hat{\mathcal{O}}_{\dot{\zeta}B}\hat{\mathcal{O}}_{AB}^{2}\hat{\mathcal{O}}_{\phi\phi}\right)\dot{\zeta}.
\end{equation}
Fortunately, the explicit expression for $\Xi$ is not needed in the following analysis. 

According to $\Xi$ is identically vanishing or not, we have two case.
If
\begin{equation}
\Xi\neq0,
\end{equation}
we have $\mathrm{rank}\bm{E}=3>\mathrm{rank}\bm{D}=2$, which implies that the set of equations for the auxiliary variables has no solution.

On the other hand, if
\begin{equation}
\Xi=0,
\end{equation}
we have $\mathrm{rank}\bm{E}=\mathrm{rank}\bm{D}=2<3$, which means
that the set of equations will be solvable. In this case, the equations
(\ref{eq:111})-(\ref{eq:113}) become
\begin{equation}
2\hat{\mathcal{O}}_{BB}B+\hat{\mathcal{O}}_{AB}A+\hat{\mathcal{O}}_{\zeta B}\zeta+\hat{\mathcal{O}}_{\dot{\zeta}B}\dot{\zeta}=0,
\end{equation}
\begin{equation}
2\hat{\mathcal{O}}_{\phi\phi}\delta\phi+\hat{\mathcal{O}}_{A\phi}A=0,
\end{equation}
from which we can solve $B$ to be
\begin{equation}
B=-\frac{\hat{\mathcal{O}}_{AB}A+\hat{\mathcal{O}}_{\zeta B}\zeta+\hat{\mathcal{O}}_{\dot{\zeta}B}\dot{\zeta}}{2\hat{\mathcal{O}}_{BB}}.\label{eq:127}
\end{equation}
Plugging the solution (\ref{eq:127}) into the last equation of motion
(\ref{eq:114}) yields the second order time derivative term of $\zeta$,
\begin{equation}
\Delta_{1}\ddot{\zeta},
\end{equation}
where
\begin{equation}
\Delta_{1}=\frac{4\hat{\mathcal{O}}_{\dot{\zeta}\dot{\zeta}}\hat{\mathcal{O}}_{BB}-\hat{\mathcal{O}}_{\dot{\zeta}B}^{2}}{2\hat{\mathcal{O}}_{BB}}.
\end{equation}
To eliminate the scalar mode, we must require
\begin{equation}
4\hat{\mathcal{O}}_{\dot{\zeta}\dot{\zeta}}\hat{\mathcal{O}}_{BB}-\hat{\mathcal{O}}_{\dot{\zeta}B}^{2}=0,
\end{equation}
which corresponds to
\begin{equation}
4a^{-2}b\left[b-3\left(c_{1}+c_{2}\right)\right]\partial^{4}=0.\label{case2eq}
\end{equation}
Recall that we have assumed $c_{1}+c_{2}\neq0$ from the beginning, and thus the only feasible solution to (\ref{case2eq}) is $c_{1}=0$. Again, this will lead to the disappearance of $\hat{K}_{ij}\hat{K}^{ij}$ and thus the kinetic term for the gravitational wave, which is unacceptable.

To conclude, there is no viable Lagrangian in Case 2.

\subsubsection{Case 3}

In this case we assume 
\begin{equation}
c_{1}+c_{2}=0,\quad-b'+b=0.
\end{equation}
As mentioned earlier, this case conflicts with the background equation of motion (\ref{bgeom_A}), and thus is unphysical.

\subsubsection{Case 4}

In this case we assume 
\begin{equation}
c_{1}+c_{2}=0,\quad d_{1}=0.
\end{equation}
The Lagrangian reduces to be
\begin{equation}
\mathcal{L}=-\frac{1}{2}bK_{ij}K^{ij}+\frac{1}{2}bK^{2}+c_{3}R+c_{4}a_{i}a^{i}+d_{2}a_{i}\mathrm{D}^{i}\phi.
\end{equation}
The operators $\hat{\mathcal{O}}_{XY}$ are given by
\begin{equation}
\hat{\mathcal{O}}_{AA}=\frac{3}{2}H^{2}\left(2b-2b'+b''\right)-c_{4}\partial^{2},\label{OAA}
\end{equation}
\begin{equation}
\hat{\mathcal{O}}_{AB}=2Ha^{-1}\left(-b'+b\right)\partial^{2},\label{OAB}
\end{equation}
\begin{equation}
\hat{\mathcal{O}}_{A\phi}=-d_{2}\partial^{2},
\end{equation}
\begin{equation}
\hat{\mathcal{O}}_{\dot{\zeta}A}=6H\left(-b+b'\right),
\end{equation}
\begin{equation}
\hat{\mathcal{O}}_{\dot{\zeta}B}=-2a^{-1}b\partial^{2},
\end{equation}
\begin{equation}
\hat{\mathcal{O}}_{\dot{\zeta}\dot{\zeta}}=3b,\label{Ozdzd}
\end{equation}
and $\hat{\mathcal{O}}_{BB}=\hat{\mathcal{O}}_{\phi\phi}=\hat{\mathcal{O}}_{\phi B}=\hat{\mathcal{O}}_{\dot{\zeta}\phi}=\hat{\mathcal{O}}_{\phi\zeta}=0$. 

Consequently, the quadratic action for the perturbation variables is
\begin{align}
S_{2}[A,B,\zeta,\delta\phi] & =\int\mathrm{d}t\mathrm{d}^{3}x\bar{N}a^{3}\left(A\hat{\mathcal{O}}_{AA}A+\zeta\hat{\mathcal{O}}_{\zeta\zeta}\zeta+A\hat{\mathcal{O}}_{AB}B+A\hat{\mathcal{O}}_{A\zeta}\zeta+A\hat{\mathcal{O}}_{A\phi}\delta\phi\right.\nonumber \\
 & \quad\left.+\zeta\hat{\mathcal{O}}_{\zeta B}B+\dot{\zeta}\hat{\mathcal{O}}_{\dot{\zeta}A}A+\dot{\zeta}\hat{\mathcal{O}}_{\dot{\zeta}B}B+\dot{\zeta}\hat{\mathcal{O}}_{\dot{\zeta}\zeta}\zeta+\dot{\zeta}\hat{\mathcal{O}}_{\dot{\zeta}\dot{\zeta}}\dot{\zeta}\right).
\end{align}
The equations of motion for the perturbation variables are
\begin{align}
2\hat{\mathcal{O}}_{AA}A+\hat{\mathcal{O}}_{AB}B+\hat{\mathcal{O}}_{A\zeta}\zeta+\hat{\mathcal{O}}_{A\phi}\delta\phi+\hat{\mathcal{O}}_{\dot{\zeta}A}\dot{\zeta} & =0,\label{eq:145}\\
\hat{\mathcal{O}}_{AB}A+\hat{\mathcal{O}}_{\zeta B}\zeta+\hat{\mathcal{O}}_{\dot{\zeta}B}\dot{\zeta} & =0,\label{eq:146}\\
\hat{\mathcal{O}}_{A\phi}A & =0,\label{eq:147}\\
\frac{1}{\bar{N}}\left[2\partial_{t}\left(a^{3}\hat{\mathcal{O}}_{\dot{\zeta}\dot{\zeta}}\dot{\zeta}\right)+\partial_{t}\left(a^{3}\hat{\mathcal{O}}_{\dot{\zeta}B}B\right)+\partial_{t}\left(a^{3}\hat{\mathcal{O}}_{\dot{\zeta}A}A\right)+\partial_{t}\left(a^{3}\hat{\mathcal{O}}_{\dot{\zeta}\zeta}\zeta\right)\right]\quad\nonumber \\
-a^{3}\left(2\hat{\mathcal{O}}_{\zeta\zeta}\zeta+\hat{\mathcal{O}}_{A\zeta}A+\hat{\mathcal{O}}_{\phi\zeta}\delta\phi+\hat{\mathcal{O}}_{\zeta B}B\right) & =0.\label{zetaeom}
\end{align}
In order to have a nontrivial solution for $A$, from (\ref{eq:147}) we must require that
\begin{equation}
\hat{\mathcal{O}}_{A\phi}=0,
\end{equation}
which implies
\begin{equation}
d_{2}=0.
\end{equation}
Then we can solve $A$ and $B$ from (\ref{eq:145})-(\ref{eq:146}) to get
\begin{equation}
A=-\frac{\hat{\mathcal{O}}_{\zeta B}\zeta+\hat{\mathcal{O}}_{\dot{\zeta}B}\dot{\zeta}}{\hat{\mathcal{O}}_{AB}},\label{Asol}
\end{equation}
and
\begin{equation}
B=\frac{\left(2\hat{\mathcal{O}}_{AA}\hat{\mathcal{O}}_{\zeta B}-\hat{\mathcal{O}}_{AB}\hat{\mathcal{O}}_{A\zeta}\right)\zeta+\left(2\hat{\mathcal{O}}_{AA}\hat{\mathcal{O}}_{\dot{\zeta}B}-\hat{\mathcal{O}}_{AB}\hat{\mathcal{O}}_{\dot{\zeta}A}\right)\dot{\zeta}}{\hat{\mathcal{O}}_{AB}^{2}},\label{Bsol}
\end{equation}
where note $\hat{\mathcal{O}}_{AB}$ given in (\ref{OAB}) is not
vanishing. 

Finally, by plugging the solutions (\ref{Asol}) and (\ref{Bsol}) into (\ref{zetaeom}), we obtain an effective equation of motion for $\zeta$ (with an undetermined $\delta\phi$), in which the second-order time derivative term can be expressed as:
\begin{equation}
\Delta_{2}\ddot{\zeta},
\end{equation}
with
\begin{equation}
\Delta_{2}=2\frac{\hat{\mathcal{O}}_{\dot{\zeta}\dot{\zeta}}\hat{\mathcal{O}}_{AB}^{2}+\hat{\mathcal{O}}_{AA}\hat{\mathcal{O}}_{\dot{\zeta}B}^{2}-\hat{\mathcal{O}}_{AB}\hat{\mathcal{O}}_{\dot{\zeta}A}\hat{\mathcal{O}}_{\dot{\zeta}B}}{\hat{\mathcal{O}}_{AB}^{2}}.
\end{equation}
Note that although $\delta\phi$ is left undetermined, it will not contribute to the $\ddot{\zeta}$ term in its equation of motion. In order to evade the unwanted scalar mode, we must impose the condition $\Delta_{2}=0$, which is equivalent to
\begin{equation}
\hat{\mathcal{O}}_{\dot{\zeta}\dot{\zeta}}\hat{\mathcal{O}}_{AB}^{2}+\hat{\mathcal{O}}_{AA}\hat{\mathcal{O}}_{\dot{\zeta}B}^{2}-\hat{\mathcal{O}}_{AB}\hat{\mathcal{O}}_{\dot{\zeta}A}\hat{\mathcal{O}}_{\dot{\zeta}B}=0.\label{cond_c4}
\end{equation}
After plugging (\ref{OAA})-(\ref{Ozdzd}) into (\ref{cond_c4}), we get 
\begin{equation}
6H^{2}a^{-2}b\left(bb''+2bb'-2b'^{2}\right)\partial^{4}-4a^{-2}b^{2}c_{4}\partial^{6}=0,
\end{equation}
which implies two equations
\begin{align}
b\left(bb''+2bb'-2b'^{2}\right) & =0,\\
b^{2}c_{4} & =0.
\end{align}
Since the background equations of motion must be satisfied, there must be $b\neq0$. The only solutions to $\Delta_{2}=0$ are thus:
\begin{equation}
	c_{4}=0,\quad b=\frac{C_{2}N}{1+C_{1}N}.
\end{equation}
As a result, in this case, the Lagrangian that does not propagate any scalar degree of freedom at the linear order is
\begin{equation}
\mathcal{L}=-\frac{1}{2}\frac{C_{2}N}{1+C_{1}N}\hat{K}_{ij}\hat{K}^{ij}+\frac{1}{3}\frac{C_{2}N}{1+C_{1}N}K^{2}+c_{3}R,\label{Lag_c4}
\end{equation}
where $c_{3}$ is a general function of $N$ and $\phi$. 

In this case, if we choose $c_{3}=1$, and let $\left|C_{2}\right|\rightarrow\infty$,
$\left|C_{1}\right|\rightarrow\infty$ while keeping $\frac{C_{2}}{C_{1}}=-2$,
the Lagrangian (\ref{Lag_c4}) reduces to be
\begin{equation}
\mathcal{L}=\hat{K}_{ij}\hat{K}^{ij}-\frac{2}{3}K^{2}+R,
\end{equation}
which is nothing but GR.

\section{Conclusion \label{sec:con}}

The spatially covariant gravity provides us with a broad playground for studying modifications of general relativity. According to the discussion in \citep{Gao:2018izs}, the spatial covariant gravity theory with non-dynamical auxiliary fields propagates three dynamical degrees of freedom, i.e., two tensorial and one scalar degrees of freedom. In light of the detection of gravitational waves, which correspond to the two tensorial degrees of freedom (TTDOFs), it is interesting to examine under which conditions the theory propagates only TTDOFs.

In this work, instead of deriving and solving the fully nonlinear TTDOF conditions as in \citep{Gao:2019twq}, we perform a perturbative analysis similar to \citep{Gao:2019lpz,Hu:2021yaq} and focus on the necessary conditions such that no scalar mode propagates at the linear order in perturbations around a cosmological background. In Sec. \ref{sec:model}, starting from the general action (\ref{action}), the background equations of motion (\ref{bgeom}) are easily obtained from the linear order action (\ref{eq:24}). These equations of motion are useful for simplifying the calculations at quadratic order. Then in Sec. \ref{sec:degen}, we derive the quadratic action for the scalar perturbations (\ref{S2}) and find that $A$, $B$, and $\delta\phi$ are auxiliary variables. By solving these auxiliary variables and deriving the effective equation of motion for the would-be dynamical perturbation variable $\zeta$, this unwanted scalar mode will be eliminated as long as condition (\ref{eq:58}) is satisfied.

As an illustration of our analysis, in Sec. \ref{sec:d2}, we consider a polynomial-type Lagrangian (\ref{eq:60}), in which all the monomials are of $d=2$ with $d$ being the total number of derivatives. Generally, if the condition (\ref{eq:57}) holds, i.e., when all the auxiliary perturbation variables $A$, $B$, $\delta\phi$ are solvable, the TTDOF condition (\ref{eq:58}) leads to the Lagrangian (\ref{Lag_OmgN0}), which propagates only TTDOFs at linear order around a cosmological background. Comparing with the result in \citep{Hu:2021yaq}, terms involving the acceleration are allowed in the Lagrangian due to the presence of the auxiliary scalar field. It is interesting that even when the condition (\ref{eq:57}) does not hold, which implies that not all of the auxiliary variables can be solved, we can still derive the degeneracy condition to eliminate the unwanted scalar mode. In this case, we obtain two viable Lagrangians (\ref{eq:103}) and (\ref{Lag_c4}). Both cases include general relativity as a special case, as long as appropriate coefficients are chosen.

We have several comments on the results presented in this work. First, the degeneracy condition derived in this work is necessary in the sense that the unwanted scalar mode is removed only at linear order in a cosmological background, which may reappear at nonlinear orders and/or around a general background. Therefore, one may perform analysis similar to those in \citep{Gao:2019lpz,Hu:2021yaq} to derive further constraints on the Lagrangian such that the scalar mode is fully removed. Second, in the example of $d=2$, the resulting Lagrangian in the case of $\varOmega\neq0$ generalizes the one derived in \citep{Hu:2021yaq}. The auxiliary scalar field $\phi$ not only affects the coefficients in the Lagrangian but also makes the acceleration term allowed. Therefore, it is interesting to explore more general cases (e.g., with $d\geq3$) where the auxiliary scalar field may play a nontrivial role, so that a novel class of TTDOF theories with such auxiliary fields can be constructed.

\begin{acknowledgments}
This work was partly supported by the Natural Science Foundation of China (NSFC) under the grant No. 11975020. 
\end{acknowledgments}

\providecommand{\href}[2]{#2}\begingroup\raggedright\endgroup


\begin{thebibliography}{10}
	
	\bibitem{Lovelock:1971yv}
	D.~Lovelock, {\it {The Einstein tensor and its generalizations}},  {\em
		J.Math.Phys.} {\bf 12} (1971) 498--501.
	
	\bibitem{Lovelock:1972vz}
	D.~Lovelock, {\it {The four-dimensionality of space and the einstein tensor}},
	{\em J. Math. Phys.} {\bf 13} (1972) 874--876.
	
	\bibitem{Afshordi:2006ad}
	N.~Afshordi, D.~J.~H. Chung, and G.~Geshnizjani, {\it {Cuscuton: A Causal Field
			Theory with an Infinite Speed of Sound}},  {\em Phys. Rev.} {\bf D75} (2007)
	083513, [\href{http://arxiv.org/abs/hep-th/0609150}{{\tt hep-th/0609150}}].
	
	\bibitem{Afshordi:2007yx}
	N.~Afshordi, D.~J.~H. Chung, M.~Doran, and G.~Geshnizjani, {\it {Cuscuton
			Cosmology: Dark Energy meets Modified Gravity}},  {\em Phys. Rev.} {\bf D75}
	(2007) 123509, [\href{http://arxiv.org/abs/astro-ph/0702002}{{\tt
			astro-ph/0702002}}].
	
	\bibitem{Mylova:2023ddj}
	M.~Mylova and N.~Afshordi, {\it {Effective Cuscuton Theory}},
	\href{http://arxiv.org/abs/2312.06066}{{\tt arXiv:2312.06066}}.
	
	\bibitem{Boruah:2017tvg}
	S.~S. Boruah, H.~J. Kim, and G.~Geshnizjani, {\it {Theory of Cosmological
			Perturbations with Cuscuton}},  {\em JCAP} {\bf 1707} (2017), no.~07 022,
	[\href{http://arxiv.org/abs/1704.01131}{{\tt arXiv:1704.01131}}].
	
	\bibitem{Boruah:2018pvq}
	S.~S. Boruah, H.~J. Kim, M.~Rouben, and G.~Geshnizjani, {\it {Cuscuton
			bounce}},  {\em JCAP} {\bf 08} (2018) 031,
	[\href{http://arxiv.org/abs/1802.06818}{{\tt arXiv:1802.06818}}].
	
	\bibitem{Bhattacharyya:2016mah}
	J.~Bhattacharyya, A.~Coates, M.~Colombo, A.~E. Gumrukcuoglu, and T.~P.
	Sotiriou, {\it {Revisiting the cuscuton as a Lorentz-violating gravity
			theory}},  {\em Phys. Rev.} {\bf D97} (2018), no.~6 064020,
	[\href{http://arxiv.org/abs/1612.01824}{{\tt arXiv:1612.01824}}].
	
	\bibitem{Quintin:2019orx}
	J.~Quintin and D.~Yoshida, {\it {Cuscuton gravity as a classically stable
			limiting curvature theory}},  {\em JCAP} {\bf 02} (2020) 016,
	[\href{http://arxiv.org/abs/1911.06040}{{\tt arXiv:1911.06040}}].
	
	\bibitem{Bartolo:2021wpt}
	N.~Bartolo, A.~Ganz, and S.~Matarrese, {\it {Cuscuton inflation}},  {\em JCAP}
	{\bf 05} (2022), no.~05 008, [\href{http://arxiv.org/abs/2111.06794}{{\tt
			arXiv:2111.06794}}].
	
	\bibitem{HosseiniMansoori:2022xnq}
	S.~A. Hosseini~Mansoori and Z.~Molaee, {\it {Multi-field Cuscuton cosmology}},
	{\em JCAP} {\bf 01} (2023) 022, [\href{http://arxiv.org/abs/2207.06720}{{\tt
			arXiv:2207.06720}}].
	
	\bibitem{Channuie:2023ddv}
	P.~Channuie, K.~Karwan, and J.~Sangtawee, {\it {Observational constraints and
			preheating in cuscuton inflation}},  {\em Eur. Phys. J. C} {\bf 83} (2023),
	no.~5 421, [\href{http://arxiv.org/abs/2301.07019}{{\tt arXiv:2301.07019}}].
	
	\bibitem{Maeda:2022ozc}
	K.-i. Maeda and S.~Panpanich, {\it {Cuscuta-Galileon cosmology: Dynamics,
			gravitational constants, and the Hubble constant}},  {\em Phys. Rev. D} {\bf
		105} (2022), no.~10 104022, [\href{http://arxiv.org/abs/2202.04908}{{\tt
			arXiv:2202.04908}}].
	
	\bibitem{Kohri:2022vst}
	K.~Kohri and K.-i. Maeda, {\it {A possible solution to the helium anomaly of
			EMPRESS VIII by cuscuton gravity theory}},  {\em PTEP} {\bf 2022} (2022),
	no.~9 091E01, [\href{http://arxiv.org/abs/2206.11257}{{\tt
			arXiv:2206.11257}}].
	
	\bibitem{Panpanich:2021lsd}
	S.~Panpanich and K.-i. Maeda, {\it {Cosmological dynamics of
			Cuscuta\textendash{}Galileon gravity}},  {\em Eur. Phys. J. C} {\bf 83}
	(2023), no.~3 240, [\href{http://arxiv.org/abs/2109.12288}{{\tt
			arXiv:2109.12288}}].
	
	\bibitem{Lima:2023str}
	F.~C.~E. Lima and C.~A.~S. Almeida, {\it {Topological solitons in the
			sigma-cuscuton model}},  {\em Eur. Phys. J. C} {\bf 83} (2023), no.~9 831,
	[\href{http://arxiv.org/abs/2301.01397}{{\tt arXiv:2301.01397}}].
	
	\bibitem{Iyonaga:2018vnu}
	A.~Iyonaga, K.~Takahashi, and T.~Kobayashi, {\it {Extended Cuscuton:
			Formulation}},  {\em JCAP} {\bf 1812} (2018), no.~12 002,
	[\href{http://arxiv.org/abs/1809.10935}{{\tt arXiv:1809.10935}}].
	
	\bibitem{Iyonaga:2020bmm}
	A.~Iyonaga, K.~Takahashi, and T.~Kobayashi, {\it {Extended Cuscuton as Dark
			Energy}},  {\em JCAP} {\bf 07} (2020) 004,
	[\href{http://arxiv.org/abs/2003.01934}{{\tt arXiv:2003.01934}}].
	
	\bibitem{Zhu:2011xe}
	T.~Zhu, Q.~Wu, A.~Wang, and F.-W. Shu, {\it {U(1) symmetry and elimination of
			spin-0 gravitons in Horava-Lifshitz gravity without the projectability
			condition}},  {\em Phys. Rev.} {\bf D84} (2011) 101502,
	[\href{http://arxiv.org/abs/1108.1237}{{\tt arXiv:1108.1237}}].
	
	\bibitem{Zhu:2011yu}
	T.~Zhu, F.-W. Shu, Q.~Wu, and A.~Wang, {\it {General covariant Horava-Lifshitz
			gravity without projectability condition and its applications to cosmology}},
	{\em Phys. Rev.} {\bf D85} (2012) 044053,
	[\href{http://arxiv.org/abs/1110.5106}{{\tt arXiv:1110.5106}}].
	
	\bibitem{Chagoya:2018yna}
	J.~Chagoya and G.~Tasinato, {\it {A new scalar-tensor realization of
			Ho\v{r}ava-Lifshitz gravity}},  \href{http://arxiv.org/abs/1805.12010}{{\tt
			arXiv:1805.12010}}.
	
	\bibitem{Afshordi:2009tt}
	N.~Afshordi, {\it {Cuscuton and low energy limit of Horava-Lifshitz gravity}},
	{\em Phys. Rev.} {\bf D80} (2009) 081502,
	[\href{http://arxiv.org/abs/0907.5201}{{\tt arXiv:0907.5201}}].
	
	\bibitem{Khoury:2011ay}
	J.~Khoury, G.~E. Miller, and A.~J. Tolley, {\it {Spatially Covariant Theories
			of a Transverse, Traceless Graviton, Part I: Formalism}},  {\em Phys.Rev.}
	{\bf D85} (2012) 084002, [\href{http://arxiv.org/abs/1108.1397}{{\tt
			arXiv:1108.1397}}].
	
	\bibitem{Khoury:2014sea}
	J.~Khoury, G.~E.~J. Miller, and A.~J. Tolley, {\it {How General Relativity and
			Lorentz Covariance Arise from the Spatially Covariant Effective Field Theory
			of the Transverse, Traceless Graviton}},  {\em Int. J. Mod. Phys. D} {\bf 23}
	(2014), no.~12 1442012, [\href{http://arxiv.org/abs/1405.5219}{{\tt
			arXiv:1405.5219}}].
	
	\bibitem{Chagoya:2016inc}
	J.~Chagoya and G.~Tasinato, {\it {A geometrical approach to degenerate
			scalar-tensor theories}},  {\em JHEP} {\bf 02} (2017) 113,
	[\href{http://arxiv.org/abs/1610.07980}{{\tt arXiv:1610.07980}}].
	
	\bibitem{Tasinato:2020fni}
	G.~Tasinato, {\it {Symmetries for scalarless scalar theories}},  {\em Phys.
		Rev. D} {\bf 102} (2020), no.~8 084009,
	[\href{http://arxiv.org/abs/2009.02157}{{\tt arXiv:2009.02157}}].
	
	\bibitem{Glavan:2019inb}
	D.~Glavan and C.~Lin, {\it {Einstein-Gauss-Bonnet Gravity in Four-Dimensional
			Spacetime}},  {\em Phys. Rev. Lett.} {\bf 124} (2020), no.~8 081301,
	[\href{http://arxiv.org/abs/1905.03601}{{\tt arXiv:1905.03601}}].
	
	\bibitem{Lin:2017oow}
	C.~Lin and S.~Mukohyama, {\it {A Class of Minimally Modified Gravity
			Theories}},  {\em JCAP} {\bf 1710} (2017), no.~10 033,
	[\href{http://arxiv.org/abs/1708.03757}{{\tt arXiv:1708.03757}}].
	
	\bibitem{Aoki:2018brq}
	K.~Aoki, A.~De~Felice, C.~Lin, S.~Mukohyama, and M.~Oliosi, {\it {Phenomenology
			in type-I minimally modified gravity}},  {\em JCAP} {\bf 1901} (2019), no.~01
	017, [\href{http://arxiv.org/abs/1810.01047}{{\tt arXiv:1810.01047}}].
	
	\bibitem{Aoki:2018zcv}
	K.~Aoki, C.~Lin, and S.~Mukohyama, {\it {Novel matter coupling in general
			relativity via canonical transformation}},  {\em Phys. Rev.} {\bf D98}
	(2018), no.~4 044022, [\href{http://arxiv.org/abs/1804.03902}{{\tt
			arXiv:1804.03902}}].
	
	\bibitem{DeFelice:2015hla}
	A.~De~Felice and S.~Mukohyama, {\it {Minimal theory of massive gravity}},  {\em
		Phys. Lett. B} {\bf 752} (2016) 302--305,
	[\href{http://arxiv.org/abs/1506.01594}{{\tt arXiv:1506.01594}}].
	
	\bibitem{DeFelice:2015moy}
	A.~De~Felice and S.~Mukohyama, {\it {Phenomenology in minimal theory of massive
			gravity}},  \href{http://arxiv.org/abs/1512.04008}{{\tt arXiv:1512.04008}}.
	
	\bibitem{Bolis:2018vzs}
	N.~Bolis, A.~De~Felice, and S.~Mukohyama, {\it {Integrated Sachs-Wolfe-galaxy
			cross-correlation bounds on the two branches of the minimal theory of massive
			gravity}},  {\em Phys. Rev. D} {\bf 98} (2018), no.~2 024010,
	[\href{http://arxiv.org/abs/1804.01790}{{\tt arXiv:1804.01790}}].
	
	\bibitem{DeFelice:2018vza}
	A.~De~Felice, F.~Larrouturou, S.~Mukohyama, and M.~Oliosi, {\it {Black holes
			and stars in the minimal theory of massive gravity}},  {\em Phys. Rev. D}
	{\bf 98} (2018), no.~10 104031, [\href{http://arxiv.org/abs/1808.01403}{{\tt
			arXiv:1808.01403}}].
	
	\bibitem{DeFelice:2021trp}
	A.~De~Felice, S.~Mukohyama, and M.~C. Pookkillath, {\it {Minimal theory of
			massive gravity and constraints on the graviton mass}},  {\em JCAP} {\bf 12}
	(2021), no.~12 011, [\href{http://arxiv.org/abs/2110.01237}{{\tt
			arXiv:2110.01237}}].
	
	\bibitem{DeFelice:2023bwq}
	A.~De~Felice, S.~Kumar, S.~Mukohyama, and R.~C. Nunes, {\it {Observational
			bounds on extended minimal theories of massive gravity: New limits on the
			graviton mass}},  \href{http://arxiv.org/abs/2311.10530}{{\tt
			arXiv:2311.10530}}.
	
	\bibitem{Mukohyama:2019unx}
	S.~Mukohyama and K.~Noui, {\it {Minimally Modified Gravity: a Hamiltonian
			Construction}},  {\em JCAP} {\bf 1907} (2019) 049,
	[\href{http://arxiv.org/abs/1905.02000}{{\tt arXiv:1905.02000}}].
	
	\bibitem{DeFelice:2020eju}
	A.~De~Felice, A.~Doll, and S.~Mukohyama, {\it {A theory of type-II minimally
			modified gravity}},  {\em JCAP} {\bf 09} (2020) 034,
	[\href{http://arxiv.org/abs/2004.12549}{{\tt arXiv:2004.12549}}].
	
	\bibitem{DeFelice:2020cpt}
	A.~De~Felice, S.~Mukohyama, and M.~C. Pookkillath, {\it {Addressing $H_0$
			tension by means of VCDM}},  {\em Phys. Lett. B} {\bf 816} (2021) 136201,
	[\href{http://arxiv.org/abs/2009.08718}{{\tt arXiv:2009.08718}}].
	
	\bibitem{Aoki:2020oqc}
	K.~Aoki, A.~De~Felice, S.~Mukohyama, K.~Noui, M.~Oliosi, and M.~C. Pookkillath,
	{\it {Minimally modified gravity fitting Planck data better than $\Lambda
			$CDM}},  {\em Eur. Phys. J. C} {\bf 80} (2020), no.~8 708,
	[\href{http://arxiv.org/abs/2005.13972}{{\tt arXiv:2005.13972}}].
	
	\bibitem{Pookkillath:2021gdp}
	M.~C. Pookkillath, {\it {Minimally Modified Gravity Fitting Planck Data Better
			Than \ensuremath{\Lambda}CDM}},  {\em Astron. Rep.} {\bf 65} (2021), no.~10
	1021--1025.
	
	\bibitem{DeFelice:2020ecp}
	A.~De~Felice, F.~Larrouturou, S.~Mukohyama, and M.~Oliosi, {\it {Minimal Theory
			of Bigravity: construction and cosmology}},  {\em JCAP} {\bf 04} (2021) 015,
	[\href{http://arxiv.org/abs/2012.01073}{{\tt arXiv:2012.01073}}].
	
	\bibitem{Aoki:2021zuy}
	K.~Aoki, F.~Di~Filippo, and S.~Mukohyama, {\it {Non-uniqueness of massless
			transverse-traceless graviton}},  {\em JCAP} {\bf 05} (2021) 071,
	[\href{http://arxiv.org/abs/2103.15044}{{\tt arXiv:2103.15044}}].
	
	\bibitem{DeFelice:2020onz}
	A.~De~Felice, A.~Doll, F.~Larrouturou, and S.~Mukohyama, {\it {Black holes in a
			type-II minimally modified gravity}},  {\em JCAP} {\bf 03} (2021) 004,
	[\href{http://arxiv.org/abs/2010.13067}{{\tt arXiv:2010.13067}}].
	
	\bibitem{DeFelice:2020prd}
	A.~De~Felice and S.~Mukohyama, {\it {Weakening gravity for dark matter in a
			type-II minimally modified gravity}},  {\em JCAP} {\bf 04} (2021) 018,
	[\href{http://arxiv.org/abs/2011.04188}{{\tt arXiv:2011.04188}}].
	
	\bibitem{DeFelice:2021xps}
	A.~De~Felice, S.~Mukohyama, and M.~C. Pookkillath, {\it {Static, spherically
			symmetric objects in type-II minimally modified gravity}},  {\em Phys. Rev.
		D} {\bf 105} (2022), no.~10 104013,
	[\href{http://arxiv.org/abs/2110.14496}{{\tt arXiv:2110.14496}}].
	
	\bibitem{DeFelice:2022uxv}
	A.~De~Felice, K.-i. Maeda, S.~Mukohyama, and M.~C. Pookkillath, {\it
		{Comparison of two theories of Type-IIa minimally modified gravity}},  {\em
		Phys. Rev. D} {\bf 106} (2022), no.~2 024028,
	[\href{http://arxiv.org/abs/2204.08294}{{\tt arXiv:2204.08294}}].
	
	\bibitem{Jalali:2023wqh}
	A.~F. Jalali, P.~Martens, and S.~Mukohyama, {\it {Spherical scalar collapse in
			a type-II minimally modified gravity}},  {\em Phys. Rev. D} {\bf 109} (2024),
	no.~4 044053, [\href{http://arxiv.org/abs/2306.10672}{{\tt
			arXiv:2306.10672}}].
	
	\bibitem{Carballo-Rubio:2018czn}
	R.~Carballo-Rubio, F.~Di~Filippo, and S.~Liberati, {\it {Minimally modified
			theories of gravity: a playground for testing the uniqueness of general
			relativity}},  {\em JCAP} {\bf 1806} (2018), no.~06 026,
	[\href{http://arxiv.org/abs/1802.02537}{{\tt arXiv:1802.02537}}]. [Erratum:
	JCAP1811,no.11,E02(2018)].
	
	\bibitem{Sangtawee:2021mhz}
	J.~Sangtawee and K.~Karwan, {\it {Inflationary model in minimally modified
			gravity theories}},  \href{http://arxiv.org/abs/2103.11463}{{\tt
			arXiv:2103.11463}}.
	
	\bibitem{Ganz:2022iiv}
	A.~Ganz, {\it {Dynamical dark energy in minimally modified gravity}},  {\em
		JCAP} {\bf 08} (2022) 074, [\href{http://arxiv.org/abs/2203.12358}{{\tt
			arXiv:2203.12358}}].
	
	\bibitem{Akarsu:2024qsi}
	O.~Akarsu, A.~De~Felice, E.~Di~Valentino, S.~Kumar, R.~C. Nunes, E.~Ozulker,
	J.~A. Vazquez, and A.~Yadav, {\it {$\Lambda_{\rm s}$CDM cosmology from a
			type-II minimally modified gravity}},
	\href{http://arxiv.org/abs/2402.07716}{{\tt arXiv:2402.07716}}.
	
	\bibitem{Yao:2020tur}
	Z.-B. Yao, M.~Oliosi, X.~Gao, and S.~Mukohyama, {\it {Minimally modified
			gravity with an auxiliary constraint: A Hamiltonian construction}},  {\em
		Phys. Rev. D} {\bf 103} (2021), no.~2 024032,
	[\href{http://arxiv.org/abs/2011.00805}{{\tt arXiv:2011.00805}}].
	
	\bibitem{Yao:2023qjd}
	Z.-B. Yao, M.~Oliosi, X.~Gao, and S.~Mukohyama, {\it {Minimally modified
			gravity with auxiliary constraints formalism}},  {\em Phys. Rev. D} {\bf 107}
	(2023), no.~10 104052, [\href{http://arxiv.org/abs/2302.02090}{{\tt
			arXiv:2302.02090}}].
	
	\bibitem{Gao:2014soa}
	X.~Gao, {\it {Unifying framework for scalar-tensor theories of gravity}},  {\em
		Phys.Rev.} {\bf D90} (2014) 081501,
	[\href{http://arxiv.org/abs/1406.0822}{{\tt arXiv:1406.0822}}].
	
	\bibitem{Gao:2014fra}
	X.~Gao, {\it {Hamiltonian analysis of spatially covariant gravity}},  {\em
		Phys.Rev.} {\bf D90} (2014), no.~10 104033,
	[\href{http://arxiv.org/abs/1409.6708}{{\tt arXiv:1409.6708}}].
	
	\bibitem{Gao:2018znj}
	X.~Gao and Z.-B. Yao, {\it {Spatially covariant gravity with velocity of the
			lapse function: the Hamiltonian analysis}},  {\em JCAP} {\bf 1905} (2019)
	024, [\href{http://arxiv.org/abs/1806.02811}{{\tt arXiv:1806.02811}}].
	
	\bibitem{Gao:2019lpz}
	X.~Gao, C.~Kang, and Z.-B. Yao, {\it {Spatially Covariant Gravity: Perturbative
			Analysis and Field Transformations}},  {\em Phys. Rev.} {\bf D99} (2019),
	no.~10 104015, [\href{http://arxiv.org/abs/1902.07702}{{\tt
			arXiv:1902.07702}}].
	
	\bibitem{Yu:2024drx}
	Y.~Yu, Z.~Chen, and X.~Gao, {\it {Spatially covariant gravity with
			nonmetricity}},  \href{http://arxiv.org/abs/2402.02565}{{\tt
			arXiv:2402.02565}}.
	
	\bibitem{Gao:2020juc}
	X.~Gao, {\it {Higher derivative scalar-tensor monomials and their
			classification}},  {\em Sci. China Phys. Mech. Astron.} {\bf 64} (2021),
	no.~1 210012, [\href{http://arxiv.org/abs/2003.11978}{{\tt
			arXiv:2003.11978}}].
	
	\bibitem{Gao:2020yzr}
	X.~Gao and Y.-M. Hu, {\it {Higher derivative scalar-tensor theory and spatially
			covariant gravity: the correspondence}},  {\em Phys. Rev. D} {\bf 102}
	(2020), no.~8 084006, [\href{http://arxiv.org/abs/2004.07752}{{\tt
			arXiv:2004.07752}}].
	
	\bibitem{Gao:2020qxy}
	X.~Gao, {\it {Higher derivative scalar-tensor theory from the spatially
			covariant gravity: a linear algebraic analysis}},  {\em JCAP} {\bf 11} (2020)
	004, [\href{http://arxiv.org/abs/2006.15633}{{\tt arXiv:2006.15633}}].
	
	\bibitem{Hu:2021bbo}
	Y.-M. Hu and X.~Gao, {\it {Covariant 3+1 correspondence of the spatially
			covariant gravity and the degeneracy conditions}},  {\em Phys. Rev. D} {\bf
		105} (2022), no.~4 044023, [\href{http://arxiv.org/abs/2111.08652}{{\tt
			arXiv:2111.08652}}].
	
	\bibitem{Horava:2008ih}
	P.~Horava, {\it {Membranes at Quantum Criticality}},  {\em JHEP} {\bf 0903}
	(2009) 020, [\href{http://arxiv.org/abs/0812.4287}{{\tt arXiv:0812.4287}}].
	
	\bibitem{Horava:2009uw}
	P.~Horava, {\it {Quantum Gravity at a Lifshitz Point}},  {\em Phys.Rev.} {\bf
		D79} (2009) 084008, [\href{http://arxiv.org/abs/0901.3775}{{\tt
			arXiv:0901.3775}}].
	
	\bibitem{Blas:2009yd}
	D.~Blas, O.~Pujolas, and S.~Sibiryakov, {\it {On the Extra Mode and
			Inconsistency of Horava Gravity}},  {\em JHEP} {\bf 0910} (2009) 029,
	[\href{http://arxiv.org/abs/0906.3046}{{\tt arXiv:0906.3046}}].
	
	\bibitem{Creminelli:2006xe}
	P.~Creminelli, M.~A. Luty, A.~Nicolis, and L.~Senatore, {\it {Starting the
			Universe: Stable Violation of the Null Energy Condition and Non-standard
			Cosmologies}},  {\em JHEP} {\bf 0612} (2006) 080,
	[\href{http://arxiv.org/abs/hep-th/0606090}{{\tt hep-th/0606090}}].
	
	\bibitem{Cheung:2007st}
	C.~Cheung, P.~Creminelli, A.~L. Fitzpatrick, J.~Kaplan, and L.~Senatore, {\it
		{The Effective Field Theory of Inflation}},  {\em JHEP} {\bf 0803} (2008)
	014, [\href{http://arxiv.org/abs/0709.0293}{{\tt arXiv:0709.0293}}].
	
	\bibitem{Creminelli:2008wc}
	P.~Creminelli, G.~D'Amico, J.~Norena, and F.~Vernizzi, {\it {The Effective
			Theory of Quintessence: the w\&lt;-1 Side Unveiled}},  {\em JCAP} {\bf 0902}
	(2009) 018, [\href{http://arxiv.org/abs/0811.0827}{{\tt arXiv:0811.0827}}].
	
	\bibitem{Gubitosi:2012hu}
	G.~Gubitosi, F.~Piazza, and F.~Vernizzi, {\it {The Effective Field Theory of
			Dark Energy}},  {\em JCAP} {\bf 1302} (2013) 032,
	[\href{http://arxiv.org/abs/1210.0201}{{\tt arXiv:1210.0201}}].
	
	\bibitem{Bloomfield:2012ff}
	J.~K. Bloomfield, E.~E. Flanagan, M.~Park, and S.~Watson, {\it {Dark energy or
			modified gravity? An effective field theory approach}},  {\em JCAP} {\bf
		1308} (2013) 010, [\href{http://arxiv.org/abs/1211.7054}{{\tt
			arXiv:1211.7054}}].
	
	\bibitem{Fujita:2015ymn}
	T.~Fujita, X.~Gao, and J.~Yokoyama, {\it {Spatially covariant theories of
			gravity: disformal transformation, cosmological perturbations and the
			Einstein frame}},  {\em JCAP} {\bf 1602} (2016), no.~02 014,
	[\href{http://arxiv.org/abs/1511.04324}{{\tt arXiv:1511.04324}}].
	
	\bibitem{Gao:2019liu}
	X.~Gao and X.-Y. Hong, {\it {Propagation of gravitational waves in a
			cosmological background}},  {\em Phys.\ Rev.\ D} {\bf 101} (2020), no.~6
	064057, [\href{http://arxiv.org/abs/1906.07131}{{\tt arXiv:1906.07131}}].
	
	\bibitem{Zhu:2022dfq}
	T.~Zhu, W.~Zhao, and A.~Wang, {\it {Polarized primordial gravitational waves in
			spatial covariant gravities}},  {\em Phys. Rev. D} {\bf 107} (2023), no.~2
	024031, [\href{http://arxiv.org/abs/2210.05259}{{\tt arXiv:2210.05259}}].
	
	\bibitem{Zhu:2022uoq}
	T.~Zhu, W.~Zhao, and A.~Wang, {\it {Gravitational wave constraints on spatial
			covariant gravities}},  {\em Phys. Rev. D} {\bf 107} (2023), no.~4 044051,
	[\href{http://arxiv.org/abs/2211.04711}{{\tt arXiv:2211.04711}}].
	
	\bibitem{Zhu:2023wci}
	T.~Zhu, W.~Zhao, J.-M. Yan, C.~Gong, and A.~Wang, {\it {Tests of modified
			gravitational wave propagations with gravitational waves}},
	\href{http://arxiv.org/abs/2304.09025}{{\tt arXiv:2304.09025}}.
	
	\bibitem{Gao:2019twq}
	X.~Gao and Z.-B. Yao, {\it {Spatially covariant gravity theories with two
			tensorial degrees of freedom: the formalism}},  {\em Phys. Rev.} {\bf D101}
	(2020), no.~6 064018, [\href{http://arxiv.org/abs/1910.13995}{{\tt
			arXiv:1910.13995}}].
	
	\bibitem{Hu:2021yaq}
	Y.-M. Hu and X.~Gao, {\it {Spatially covariant gravity with 2 degrees of
			freedom: Perturbative analysis}},  {\em Phys. Rev. D} {\bf 104} (2021),
	no.~10 104007, [\href{http://arxiv.org/abs/2104.07615}{{\tt
			arXiv:2104.07615}}].
	
	\bibitem{Lin:2020nro}
	J.~Lin, Y.~Gong, Y.~Lu, and F.~Zhang, {\it {Spatially covariant gravity with a
			dynamic lapse function}},  {\em Phys. Rev. D} {\bf 103} (2021), no.~6 064020,
	[\href{http://arxiv.org/abs/2011.05739}{{\tt arXiv:2011.05739}}].
	
	\bibitem{Iyonaga:2021yfv}
	A.~Iyonaga and T.~Kobayashi, {\it {Distinguishing modified gravity with just
			two tensorial degrees of freedom from general relativity: Black holes,
			cosmology, and matter coupling}},  {\em Phys. Rev. D} {\bf 104} (2021),
	no.~12 124020, [\href{http://arxiv.org/abs/2109.10615}{{\tt
			arXiv:2109.10615}}].
	
	\bibitem{Hiramatsu:2022ahs}
	T.~Hiramatsu and T.~Kobayashi, {\it {Testing gravity with the cosmic microwave
			background: constraints on modified gravity with two tensorial degrees of
			freedom}},  {\em JCAP} {\bf 07} (2022), no.~07 040,
	[\href{http://arxiv.org/abs/2205.04688}{{\tt arXiv:2205.04688}}].
	
	\bibitem{Saito:2023bhn}
	J.~Saito and T.~Kobayashi, {\it {Black hole perturbations in spatially
			covariant gravity with just two tensorial degrees of freedom}},  {\em Phys.
		Rev. D} {\bf 108} (2023), no.~10 104063,
	[\href{http://arxiv.org/abs/2308.00267}{{\tt arXiv:2308.00267}}].
	
	\bibitem{Chakraborty:2023jek}
	S.~Chakraborty, K.~Karwan, and J.~Sangtawee, {\it {Observational predictions of
			inflationary model in spatially covariant gravity with two tensorial degrees
			of freedom for gravity}},  \href{http://arxiv.org/abs/2308.09508}{{\tt
			arXiv:2308.09508}}.
	
	\bibitem{Gao:2018izs}
	X.~Gao, M.~Yamaguchi, and D.~Yoshida, {\it {Higher derivative scalar-tensor
			theory through a non-dynamical scalar field}},  {\em JCAP} {\bf 1903} (2019)
	006, [\href{http://arxiv.org/abs/1810.07434}{{\tt arXiv:1810.07434}}].
	
\end{thebibliography}
\end{document}